\documentclass[conference]{IEEEtran}
\IEEEoverridecommandlockouts

\usepackage{cite}
\usepackage{amsmath,amssymb,amsfonts}
\usepackage{algorithmic}
\usepackage{graphicx}
\usepackage{textcomp}
\usepackage{changepage}
\usepackage{tabularx}
\usepackage{comment}  
\usepackage{algorithm}
\usepackage{subcaption}
\usepackage{lipsum}
\usepackage{caption}
\usepackage{xspace}

\usepackage[letterpaper, top=.75in, bottom=1in, left=0.625in, right=0.625in]{geometry}
\usepackage{xcolor}
 \def\BibTeX{{\rm B\kern-.05em{\sc i\kern-.025em b}\kern-.08em
    T\kern-.1667em\lower.7ex\hbox{E}\kern-.125emX}}

\begin{document}

\title{Digital Twin-Assisted Fast Beam Alignment for mmWave MIMO Networks}
\author{Muhammad Kamran Saeed and Ashfaq Khokhar,~\IEEEmembership{Fellow,~IEEE}
\vspace*{-0.80 cm}
\thanks{This research was partly funded by Palmer Department Chair Endowment Funds at Iowa State University.}
\thanks{M. K. Saeed is with the Dept. of Electrical and Computer Engineering, Iowa State University, Ames, IA, USA (email: muhkamransaeed@gmail.com). A. Khokhar is with the College of Engineering, Kansas State University, Manhattan, KS, USA (email: akhokhar@k-state.edu).}
}
\mark{IEEEE TRANSACTIONS ON NETWORK AND SERVICE MANAGEMENT}

{}

\maketitle

\begin{abstract}
Millimeter-wave (mmWave) communication relies on directional beam sweeping for initial access, which incurs substantial latency and signaling overhead as antenna arrays grow large and deployments become denser. Machine-learning-based beam prediction can reduce this overhead, but its effectiveness depends on training data that accurately reflects the deployment environment, a requirement that is difficult to meet in practice. This paper presents a latency-aware beam management architecture that integrates digital-twin-assisted prediction, region-based localization, weighted ensemble learning, and explainability-driven beam pruning into a unified framework. A digital twin (DT) generates site-specific ray-traced channel knowledge offline, enabling environment-aware beam prediction without exhaustive real-world measurements and supporting both initial access and beam tracking. The proposed ensemble learning scheme combines a base classifier and a specialist classifier through reliability-weighted fusion to improve prediction accuracy and robustness, while explainable artificial intelligence (XAI)-based attribution pruning reduces the beam codebook to a small subset of the most informative beams for beam allocation. To capture the distributional mismatch between the DT and real-world environments: the network was trained on imperfect digital-twin data but evaluated on real-world surrobase data. Results show that the proposed method achieves 96.80\% top-1 and 98.36\% top-3 beam prediction accuracy, while reducing over-the-air sounding overhead by nearly an order of magnitude relative to exhaustive sweeping, outperforming convolutional and deep neural network baselines from prior work.
\end{abstract}


\section{Introduction}
Millimeter-wave (mmWave) communication is a key enabler of next-generation wireless systems, owing to the large available bandwidth that supports extremely high data rates and low-latency transmission. The short wavelength at mmWave frequencies further enables the use of large-scale antenna arrays, allowing highly directional beamforming that enhances spatial reuse and interference management in dense deployments.
Despite these advantages, mmWave communications face significant challenges stemming from severe path loss and high sensitivity to blockage, which complicate reliable coverage and link maintenance, particularly in dynamic and non-line-of-sight
(NLOS) environments \cite{giordani2019tutorial}. To address these issues, current mmWave systems rely on directional beam sweeping for initial access (IA). In this process, the base station periodically transmits synchronization signals over a predefined set of beams, while the user equipment scans its receive beams to detect and evaluate potential links. Once a suitable transmit--receive beam pair is identified, beam refinement and random access signaling are performed to establish the connection. In 5G new radio (NR), IA is realized through synchronization signal blocks (SSBs) mapped to spatial beams. Up to 64 SSBs may be transmitted within a single beam-sweeping burst spanning a 5~ms window, with a default periodicity of 20~ms. Because each SSB corresponds to a distinct beam, the user equipment (UE) may need to wait for an entire burst to assess all candidates, and must re-scan whenever mobility or blockage disrupts beam alignment, which occurs frequently in dynamic environments. Consequently, IA can incur considerable latency and overhead, especially in dense networks, high-mobility scenarios, and systems with large antenna arrays, where the channel coherence time may be shorter than the SSB periodicity~\cite{giordani2019tutorial,xue2024survey}.

The IA procedure can be viewed as comprising two core components. The first is beam sweeping and measurement, in which synchronization signals are transmitted across multiple beams and the receiver measures the resulting signal strengths. The second is beam selection (or prediction), in which these measurements are processed to determine the optimal transmit--receive beam pair for link establishment, with beam prediction guiding the final access decision \cite{xue2024aiml}. Among these two stages, beam sweeping and measurement accounts for most of the access delay because the total duration increases linearly with the number of beams that must be scanned. To address this bottleneck, machine learning (ML)--based beam prediction reduces beam training overhead by learning the relationship between partial measurements or side information and optimal beam directions, thereby limiting the need for exhaustive beam sweeping during initial access and beam tracking \cite{jiang2023digital,xue2024aiml,ioannou2026}. Prior studies show that ML models can accurately identify candidate beams using only a small subset of sensing beams or other low-dimensional observations, significantly shortening training time while maintaining near-optimal performance \cite{xue2024aiml,khan2025explainable}. Consequently, ML-driven beam management is increasingly recognized as a key enabler for reducing signaling overhead and latency in mmWave and 5G-Advanced systems \cite{xue2024survey}.

Although ML--based beam prediction has shown strong potential for reducing beam training overhead, its effectiveness depends heavily on how well the training data reflects the target deployment environment. Prior studies indicate that models trained on synthetic or limited datasets can experience noticeable performance degradation in real-world scenarios because the underlying data distribution may not accurately capture actual propagation conditions \cite{jiang2023digital,khan2026digital}. Therefore, reliable ML-based beam prediction requires either training data that closely matches the deployment setting or adaptation mechanisms that can incorporate limited real-world measurements. One promising way to address this challenge is through DT--based approaches, which use virtual replicas of the physical environment to generate site-specific channel and beam information offline. By producing ray-tracing data that closely reflects the characteristics of the target environment, digital twins (DTs) enable beam prediction models to be trained on more representative data and to infer suitable beams directly, thereby substantially reducing the need for sequential beam sweeping during initial access and tracking \cite{jiang2023digital}.

DT--based beam management is highly effective when the virtual replica accurately captures the geometry, materials, and dynamics of the physical environment, thereby enabling reliable beam prediction with minimal real-world interaction \cite{jiang2023digital,zhang2023digital, zhang2026}. In practice, however, achieving such a near-perfect DT is challenging, since modeling errors, environmental dynamics, and incomplete information inevitably create mismatches between the digital and physical worlds \cite{jiang2023digital,khan2026digital}. These mismatches can undermine the robustness of digital-twin-aided beam prediction, making mechanisms such as model adaptation and robustness enhancement essential for mitigating performance degradation \cite{khan2026digital,zhang2023digital}. To address this limitation, we propose a region-based beam training framework that partitions the coverage area into localized regions. Rather than performing beam training across the entire spatial domain, the proposed method restricts beam exploration to region-specific subsets, thereby confining the impact of digital-twin inaccuracies to local areas. This regionalization supports more adaptive and robust beam training by accommodating localized mismatches between the DT and the physical environment, while also reducing unnecessary training overhead. To further reduce beam training overhead, we incorporate explainable artificial intelligence (XAI) into the beam management framework to enable more selective and informed beam exploration instead of exhaustive sweeping. Recent studies indicate that XAI can identify the most informative sensing beams or spatial directions, allowing redundant measurements to be pruned while preserving reliable beam alignment performance \cite{khan2026digital, khan2024trustworthy}.

In this paper, we propose a latency-aware beam management architecture that combines four complementary components to reduce beam training overhead while maintaining reliable beam alignment. First, a DT generates site-specific channel knowledge that supports environment-aware beam prediction without requiring exhaustive real-world measurements. Second, to mitigate the impact of digital-twin imperfections, a region-based framework partitions the coverage area into localized regions and confines beam training within each region, thereby limiting the propagation of digital-physical mismatch across the network. Third, within each region, a base--specialist classifier adapts inference cost to sample difficulty: a lightweight base resolves confident cases on its own, while uncertain samples are escalated to a specialist and fused with the base through a reliability-weighted ensemble that combines complementary base predictors to improve accuracy beyond any individual classifier and strengthen robustness under noisy, mismatched digital-twin conditions. Fourth, XAI is integrated into the regional beam training process to identify and prioritize the most informative beams within each region, enabling targeted exploration and further reducing training overhead. By jointly combining digital-twin-assisted prediction, region-based localization, ensemble-based regional prediction, and XAI-guided beam prioritization, the proposed framework improves robustness to environmental mismatch and provides an efficient, scalable solution for low-latency beam management in mmWave systems.

\subsection{Literature Review}
\subsubsection{Learning-Based Beam Prediction from In-Band Measurements}
A large body of work reduces beam-training overhead by inferring narrow beams from a small set of measurements. Ma \textit{et al.}~\cite{ma2021deep} map coarse wide-beam measurements to refined narrow beams through data-driven calibration, avoiding exhaustive codebook scanning while preserving alignment accuracy. Liu \textit{et al.}~\cite{liu2023deep} extend this idea to near-field extremely large-scale multiple-input multiple-output (XL-MIMO), mapping far-field wide-beam measurements to near-field beam decisions without exhaustive near-field codebook search. Deng \textit{et al.}~\cite{Deng2025} instead exploit sub-6~GHz channels, predicting the mmWave beamspace with a convolutional neural network (CNN) and performing hybrid beamforming via a multi-layer graph neural network (GNN) and linear minimum mean-square error (LMMSE) combining. Chen \textit{et al.}~\cite{chen2024beam} propose a unified XL-MIMO framework that integrates beam training, refinement, and tracking across near- and far-field regimes under practical hybrid architectures. These methods, however, require measurements for every prediction and train a single model over the entire coverage area.

\subsubsection{Position- and Sensing-Aided Beam Prediction}
Side information acquired outside the communication band offers another route to reducing beam-training overhead. Heng \textit{et al.}~\cite{Heng2021} predict the optimal access point (AP) and beam directly from user equipment global positioning system (GPS) coordinates, shrinking the joint AP--beam search space on ray-traced urban data. Nugroho \textit{et al.}~\cite{Nugroho2025} extend position-aided prediction to unmanned aerial vehicle (UAV) links, jointly forecasting current and future beams from sequential position data. Mollah \textit{et al.}~\cite{Mollah2026} fuse multi-modal out-of-band sensing data to predict the optimal beam index in vehicle-to-infrastructure (V2I) and vehicle-to-vehicle (V2V) links, while Rasheed \textit{et al.}~\cite{Rasheed2025} combine transformer-based trajectory prediction with federated multi-agent deep reinforcement learning (DRL) to sustain alignment under high mobility. However, Morais \textit{et al.}~\cite{Morais2023} show that position-aided models validated on synthetic data degrade under real commercial off-the-shelf (COTS) GPS measurements, exposing a gap between simulated and deployed performance.

\subsubsection{Environment- and Region-Aware Beam Management}
Another line of work exploits spatial structure to make beam training more selective. Zhang \textit{et al.}~\cite{zhang2024environment} leverage the spatial non-stationarity of near-field XL-MIMO channels through a two-stage strategy that first identifies the user's visibility region and associated subarray, then performs fine alignment within that region using a learned hierarchical codebook. Alawadi \textit{et al.}~\cite{alawadi2025hotspot} predict emerging traffic hotspots with multiple learning models and proactively steer beams toward anticipated high-demand areas. In these works, regional context guides the beam search but is not used to specialize the predictor to local propagation conditions.

\subsubsection{DT-Aided Beam Management}
DTs reduce the cost of acquiring real-world training data. Zeulin \textit{et al.}~\cite{Zeulin2022} train a lightweight ML beam predictor for time-sensitive industrial Internet of Things (IoT) using a map-based channel model derived from a real-world-calibrated, ray-traced 3D environment, narrowing the beam search space and reducing training overhead. Jiang \textit{et al.}~\cite{jiang2023digital} use 3D environment models and ray tracing to generate synthetic channel data for training beam predictors, showing in a position-aided mmWave case study that DT-trained models achieve near-optimal performance while reducing real-world beam sweeping. Alkhateeb \textit{et al.}~\cite{Alkhateeb2025} learn site-specific line-of-sight (LOS) and NLOS beam codebooks from DT-generated channels without costly real-world interaction, and identify ray-tracing accuracy as the dominant factor governing DT codebook performance. Zhang \textit{et al.}~\cite{zhang2023digital} use virtual environment replicas for sample-efficient beam pattern design for interference nulling without explicit channel state information (CSI), extending the role of DTs beyond beam prediction. Bonora \textit{et al.}~\cite{bonora2026digital} apply DT-guided iterative beamforming optimization to interference mitigation in amplify-and-forward relay-assisted MIMO systems and validate it on a custom mmWave testbed. Despite this progress, how DT fidelity affects beam prediction accuracy remains largely unexplored, even though simulation--reality mismatch can substantially degrade performance~\cite{Morais2023}.

\subsubsection{Explainable AI for Beam Management}
Explainable artificial intelligence (XAI) has recently moved from interpreting models to actively reducing overhead. Khan \textit{et al.}~\cite{khan2025explainable, khan2026digital} use XAI to identify the most informative sensing beams and prune redundant measurements without degrading alignment reliability. In a vehicle-to-everything (V2X) setting, Khan \textit{et al.}~\cite{khan2025explainableV2X} use SHapley Additive exPlanations (SHAP)-based feature importance to shrink the state space and model complexity of multi-agent DRL with minimal performance loss. These studies apply XAI-guided reduction to a single global model and do not exploit region-specific specialization.

\subsubsection{Summary and Positioning}
In summary, existing approaches reduce beam-training overhead through measurement-to-beam mapping, out-of-band side information, regional context, DT-generated data, or XAI-guided pruning, but largely in isolation, with a single global model, and with limited attention to DT fidelity and inference complexity. This paper unifies these directions in a DT-enabled, XAI-guided cascade framework that (i) partitions the coverage area into hexagonal regions using primary and auxiliary tilings, (ii) routes samples to a lightweight base model or a specialist predictor based on prediction confidence, with reliability-weighted ensemble fusion, (iii) constructs region-specific beam subsets via TreeSHAP-based pruning, and (iv) quantifies how DT fidelity, modeled by the number of retained propagation paths, affects prediction accuracy and computational cost.

\subsection{Contributions}

The main contributions of this paper are summarized as follows:

\begin{itemize}

     \item We propose a latency-aware beam management architecture that leverages a DT to generate site-specific channel knowledge and support environment-aware beam prediction, thereby reducing the need for exhaustive beam sweeping during initial access and beam tracking.
    
    \item We develop a region-based beam training framework that partitions the coverage area into localized regions and restricts beam exploration to the relevant region, mitigating digital-twin mismatch and reducing unnecessary global training overhead.

     \item We design a cascaded base--specialist ensemble learning architecture that combines two complementary base predictors with confidence-based routing and reliability-weighted fusion: confident samples are resolved by the lightweight base alone, while uncertain samples are jointly predicted by the base and specialist with model-specific weights assigned according to their reliability. This design improves beam prediction accuracy beyond what any individual classifier achieves and strengthens the robustness of the regional beam selection stage under noisy and mismatched digital-twin conditions.
    
     \item We integrate XAI into the regional beam management process to identify and prioritize the most informative sensing beams within each region, enabling targeted beam exploration and further reducing training overhead while preserving reliable beam alignment.
     
    \item By jointly combining digital-twin-assisted prediction, regional localization, and XAI-guided beam prioritization, the proposed framework provides a robust and scalable solution for low-latency beam management in practical mmWave systems where the DT is imperfect and real-world interaction must be minimized.

\end{itemize}

\begin{figure}[t]
    \centering
    \vspace{-3mm}
    \includegraphics[width=9cm]{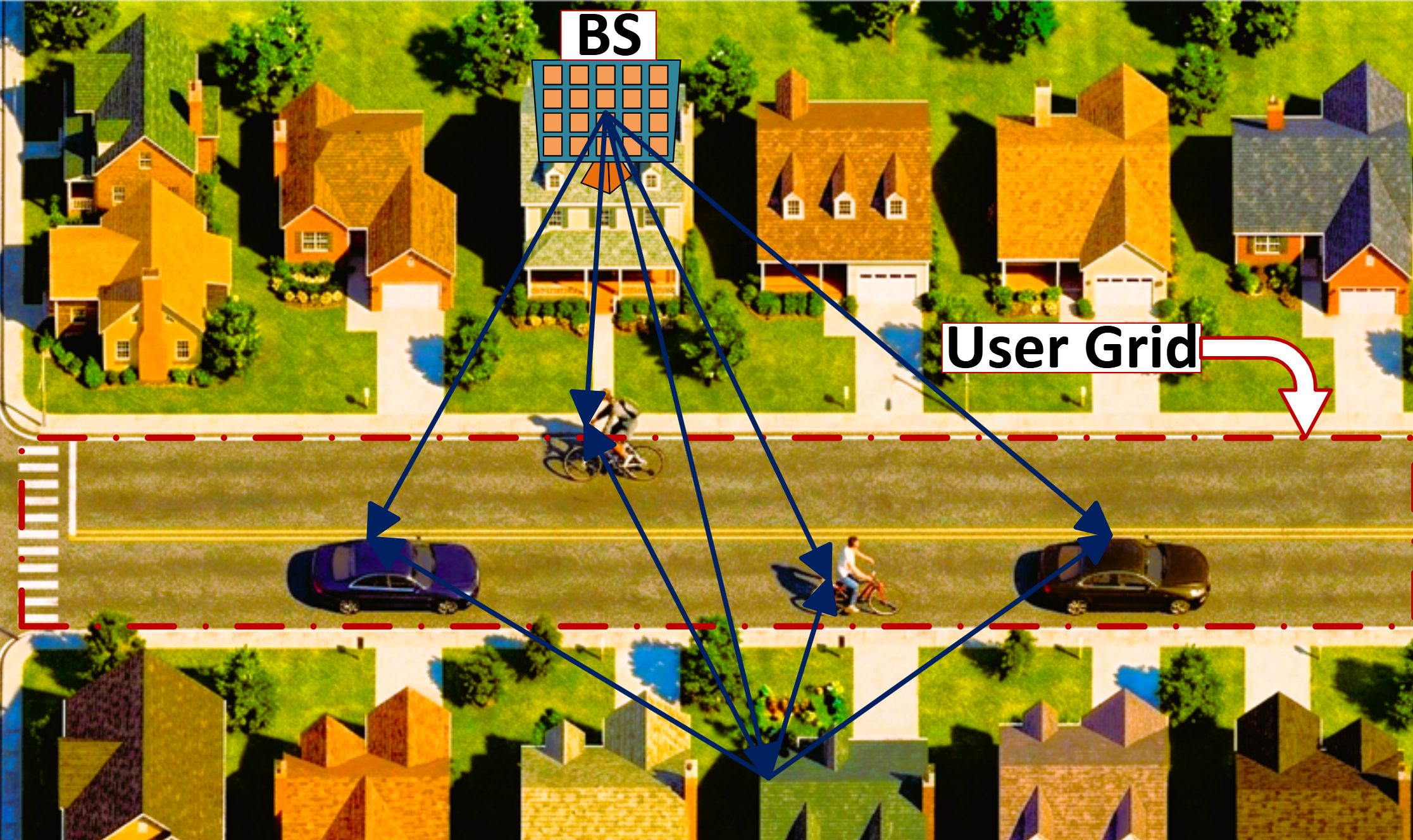}
    \caption{Downlink of a mmWave system with active users in the user grid.}
    \label{fig:1}
    \vspace{-5mm}
\end{figure}

\section{System Model}
\label{sec:system_model}
 
We consider the downlink of a mmWave cellular system in which a single base station (BS), equipped with a uniform planar array (UPA) of $N$ antenna elements, serves a set of $K$ single-antenna user equipments (UEs) indexed by $\mathcal{K} = \{1, 2, \ldots, K\}$. The system operates at carrier frequency $f_c$ with signal bandwidth $W$, and since mmWave deployments typically satisfy $W \ll f_c$, the array steering response is treated as frequency-flat across the operating band; this assumption is standard in the mmWave beam management literature and allows the precoder to be designed at the carrier wavelength without incurring beam-squint distortion. Owing to the high path loss, limited scattering, and quasi-optical propagation characteristics of the mmWave band, the BS-UE channel exhibits pronounced spatial sparsity and is most accurately captured by a geometric ray-based representation. Accordingly, the downlink channel vector $\mathbf{g}_k \in \mathbb{C}^{N \times 1}$ from the BS to the $k$-th UE is modeled as a superposition of $L_k$ dominant propagation paths, i.e.,
\begin{equation}
\mathbf{g}_k = \sqrt{\frac{N}{L_k}} \sum_{l=1}^{L_k} \beta_{k,l}\, \mathbf{a}\!\left(\phi_{k,l},\, \psi_{k,l}\right),
\label{eq:channel}
\end{equation}
where $\beta_{k,l} \in \mathbb{C}$ denotes the complex-valued small-scale fading coefficient of the $l$-th ray, while $\phi_{k,l} \in [-\pi, \pi]$ and $\psi_{k,l} \in [-\pi/2, \pi/2]$ represent its azimuth and elevation angles of departure (AoDs) observed at the BS, respectively.
The transmit array steering vector $\mathbf{a}(\phi, \psi) \in \mathbb{C}^{N \times 1}$ associated with AoD pair $(\phi, \psi)$ depends on the geometric layout of the UPA. Assuming $N = N_h N_v$ elements arranged in $N_h$ columns and $N_v$ rows with horizontal and vertical inter-element spacings $d_H$ and $d_V$, the steering vector is given by
\begin{equation}
\mathbf{a}(\phi, \psi) = \frac{1}{\sqrt{N}} \left[ \mathrm{e}^{j \tilde{\kappa} \, \xi_1(\phi,\psi)},\, \ldots,\, \mathrm{e}^{j \tilde{\kappa} \, \xi_N(\phi,\psi)} \right]^{\mathsf{T}},
\label{eq:steering}
\end{equation}
where $\tilde{\kappa} = 2\pi/\lambda_c$ denotes the wavenumber corresponding to carrier wavelength $\lambda_c$, and the spatial phase contribution of the $n$-th antenna element is
\begin{equation}
\xi_n(\phi, \psi) = \alpha_n d_H \sin(\phi)\cos(\psi) + \gamma_n d_V \sin(\psi),
\label{eq:phase_term}
\end{equation}
with $(\alpha_n, \gamma_n)$ denoting the horizontal and vertical position indices of element $n$ within the planar grid are spaced half-wavelength, i.e., $d_H = d_V = \lambda_c / 2$. To strike a favorable trade-off between hardware cost, power dissipation, and beamforming flexibility, the BS adopts an analog-only transmit architecture in which all $N$ antenna elements are driven by a single radio-frequency (RF) chain through a network of digitally controlled phase shifters. Under this front-end, the precoder $\mathbf{v} \in \mathbb{C}^{N \times 1}$ is constrained to have unit-modulus entries and is expressed as,
\begin{equation}
\mathbf{v} = \frac{1}{\sqrt{N}} \left[ e^{j\theta_1},\, e^{j\theta_2},\, \ldots,\, e^{j\theta_N} \right]^{\mathsf{T}},
\label{eq:precoder}
\end{equation}
where $\theta_n \in [0, 2\pi)$ is the phase shift introduced at the $n$-th antenna port. The normalization $1/\sqrt{N}$ ensures the unit-norm constraint $\|\mathbf{v}\|^2 = 1$, thereby fixing the radiated power independently of the chosen beam direction. Because optimizing the continuous phase variables $\{\theta_n\}$ over $\mathbb{R}^N$ is computationally prohibitive and incompatible with practical initial-access protocols (e.g., 5G NR SSB-based beam sweeping), the BS instead employs a finite-cardinality beam codebook of pre-designed analog beamformers,
\begin{equation}
\mathcal{V} = \left\{ \mathbf{v}_1,\, \mathbf{v}_2,\, \ldots,\, \mathbf{v}_C \right\}, \qquad \mathbf{v}_c \in \mathbb{C}^{N \times 1},
\end{equation}
where $C$ denotes the codebook size. Every codeword inherits the constant-modulus constraint imposed by the analog architecture, namely
\begin{equation}
\|\mathbf{v}_c\|^2 = 1, \qquad \forall c \in \{1, 2, \ldots, C\}.
\label{eq:cm_constraint}
\end{equation}
The codebook may equivalently be stacked into matrix form as $\mathcal{V}  = [\mathbf{v}_1, \mathbf{v}_2, \ldots, \mathbf{v}_C] \in \mathbb{C}^{N \times C}$, which serves as the sweeping dictionary used during beam training. Common constructions include DFT, oversampled DFT (O-DFT), and hierarchical multi-resolution codebooks. DFT codebooks give broad coverage with $C=N$ orthogonal beams but lack the resolution needed for precise alignment in narrow-beam, high-frequency regimes; O-DFT codebooks improve resolution at the cost of higher training overhead from evaluating more beams. The proposed framework is agnostic to codebook design, requiring only that $\mathcal{V}$ satisfy \eqref{eq:cm_constraint}. During beam training, the BS sequentially probes the angular domain by transmitting known pilot symbols through each codeword in $\mathcal{V}$. When the $c$-th codeword $\mathbf{v}_c$ is activated, the baseband signal observed at the $k$-th UE is
\begin{equation}
\overline{y}_k^{(c)} = \sqrt{\varrho}\, \mathbf{g}_k^{\mathsf{H}} \mathbf{v}_c\, \overline{x} + z_k,
\label{eq:rx_signal}
\end{equation}
where $\varrho$ is the BS transmit power, $\overline{x} \in \mathbb{C}$ denotes the unit-energy pilot symbol with $\mathbb{E}\{|\overline{x}|^2\} = 1$, and $z_k \sim \mathcal{CN}(0, \sigma^2)$ models the additive white Gaussian noise (AWGN) at the UE receiver with variance $\sigma^2$. The corresponding instantaneous received signal-to-noise ratio (SNR) at user $k$ under beam $\mathbf{v}_c$ is therefore
\begin{equation}
\gamma_k^{(c)} = \frac{\varrho \left| \mathbf{g}_k^{\mathsf{H}} \mathbf{v}_c \right|^2}{\sigma^2}.
\label{eq:snr}
\end{equation}

\section{Problem Formulation}
\label{sec:problem_formulation}
 
The system model introduced in Section~\ref{sec:system_model} reduces, at its core, to the per-user beam selection problem
\begin{equation}
c_k^{\star} = \arg\max_{c \in \{1, 2, \ldots, C\}} \left| \mathbf{g}_k^{\mathsf{H}} \mathbf{v}_c \right|^2,
\qquad \forall k \in \mathcal{K}.
\label{eq:opt_beam}
\end{equation}
Solving \eqref{eq:opt_beam} requires evaluating the effective beamforming gain $|\mathbf{g}_k^{\mathsf{H}} \mathbf{v}_c|^2$ over the codebook $\mathcal{V}$ for every active user. In practice, however, the instantaneous channel $\mathbf{g}_k$ is not available a priori, since acquiring it would itself demand a full sweep across $\mathcal{V}$. This circular dependency between channel knowledge and beam selection lies at the heart of the mmWave beam management problem.
 
\subsection{Limitations of Exhaustive Beam Sweeping}
 
The conventional approach, exhaustive beam sweeping (EBS), has the BS sequentially probe every codeword in $\mathcal{V}$, with the UE measuring received signal strength per beam and reporting the strongest index. While EBS guarantees identification of $c_k^{\star}$ in static channels, its overhead scales with $C$, which can approach or exceed the channel coherence interval in dynamic environments, causing outdated beam selections and conflicting with the low-latency demands of 5G NR and 6G. It also wastes energy and feedback overhead by activating the RF chain across all $C$ codewords, even though only a few beams typically offer meaningful gain. These limitations motivate approaches that exploit prior environmental information to prune the search space before over-the-air measurement.
\subsection{Site-Specific Prior via a DT}
 
In typical deployments, the propagation environment surrounding the BS, including building geometry, material properties, street layouts, and reflector locations, evolves on time scales far slower than the user channels themselves. This observation enables the construction of a DT: a site-specific computational replica of the physical environment from which synthetic channel realizations $\hat{\mathbf{g}}_k$ can be generated, for instance via ray-tracing over a three-dimensional model of the cell. Let $\hat{\mathcal{V}}_k \subseteq \mathcal{V}$ denote a candidate beam subset predicted from DT-derived priors for user $k$. The beam selection task can then be relaxed from \eqref{eq:opt_beam} to the reduced-search problem
\begin{equation}
\hat{c}_k = \arg\max_{c \,:\, \mathbf{v}_c \in \hat{\mathcal{V}}_k} \left| \mathbf{g}_k^{\mathsf{H}} \mathbf{v}_c \right|^2,
\label{eq:dt_beam}
\end{equation}
with overhead $|\hat{\mathcal{V}}_k| \ll C$. The fidelity of \eqref{eq:dt_beam} hinges on the agreement between the DT and the true environment, captured by a model-mismatch error
\begin{equation}
\Delta_k = \left\| \mathbf{g}_k - \hat{\mathbf{g}}_k \right\|^2,
\label{eq:mismatch}
\end{equation}
which is generally non-zero due to dynamic obstacles, calibration drift, and unmodeled scatterers. When $\Delta_k$ is large, $\hat{\mathcal{V}}_k$ may fail to contain $\mathbf{v}_{c_k^{\star}}$, leading to a beam-misalignment penalty
\begin{equation}
\eta_k = \left| \mathbf{g}_k^{\mathsf{H}} \mathbf{v}_{c_k^{\star}} \right|^2 - \left| \mathbf{g}_k^{\mathsf{H}} \mathbf{v}_{\hat{c}_k} \right|^2 \geq 0.
\label{eq:misalign_penalty}
\end{equation}
A central challenge is therefore to design a beam management framework that exploits DT priors aggressively enough to shrink $|\hat{\mathcal{V}}_k|$, yet remains resilient when $\Delta_k$ is non-negligible.
 
\subsection{Problem Statement}
 
Let $\mathcal{T}_{\mathrm{train}}$ denote the total beam-training time consumed for serving all $K$ users in a given training cycle, and let $\eta_k$ be the per-user misalignment penalty in \eqref{eq:misalign_penalty}. We formulate the latency-aware beam management problem as
\begin{align}
\mathop{\mathrm{minimize}}_{\{\hat{\mathcal{V}}_k\}_{k \in \mathcal{K}}}
\quad & \mathcal{T}_{\mathrm{train}}\left(\{\hat{\mathcal{V}}_k\}\right)
\label{eq:problem_obj}\\
\mathrm{subject\ to} \quad
& \Pr\!\left( \eta_k \leq \epsilon \right) \geq 1 - \delta,
\quad \forall k \in \mathcal{K},
\label{eq:problem_reliability}\\
& \hat{\mathcal{V}}_k \subseteq \mathcal{V},
\quad \forall k \in \mathcal{K},
\label{eq:problem_subset}
\end{align}
where $\epsilon > 0$ specifies the maximum tolerable beamforming gain loss, and $\delta \in (0,1)$ controls the outage probability associated with that loss. The constraint \eqref{eq:problem_reliability} ensures that the proposed framework retains alignment reliability even under DT mismatch, while \eqref{eq:problem_obj} captures the overarching goal of minimizing training latency.
 
The optimization in \eqref{eq:problem_obj}--\eqref{eq:problem_subset} is intractable in its general form, for three reasons. First, the true channel $\mathbf{g}_k$ governing $\eta_k$ is unknown at the time the subset $\hat{\mathcal{V}}_k$ must be committed to. Second, the relationship between DT-generated priors and the optimal beam $c_k^{\star}$ is highly non-linear and environment-dependent, precluding closed-form solutions. Third, even if a data-driven predictor for $\hat{\mathcal{V}}_k$ were available, its black-box nature offers no principled mechanism for further pruning redundant beams within a candidate subset.
 
\subsection{Design Objectives of the Proposed Framework}
 
The above difficulties motivate three coupled design objectives that jointly guide the framework developed in the remainder of this paper. First, site-specific prediction exploits DT-generated channel knowledge to construct environment-aware beam predictions $\hat{\mathcal{V}}_k$, substantially reducing the candidate set relative to the full codebook $\mathcal{V}$ and avoiding exhaustive real-world measurements. Second, robustness to DT mismatch is achieved through regionalization, where the coverage area is partitioned into localized regions $\{\mathcal{R}_1, \mathcal{R}_2, \ldots, \mathcal{R}_R\}$ and beam exploration for each user is restricted to its corresponding region. This localized scope limits the impact of DT inaccuracies to the affected region and improves resilience to local environmental discrepancies. Third, interpretable beam-level pruning uses XAI to attribute prediction confidence to individual sensing beams, rank codewords according to their relevance, and retain only those whose contribution exceeds a learned threshold. This further reduces the regional candidate set $\hat{\mathcal{V}}_k$ to an importance-ordered probing sequence, enabling targeted exploration while reducing $\mathcal{T}_{\mathrm{train}}$ and maintaining reliable beam alignment. Collectively, these objectives translate the optimization in \eqref{eq:problem_obj}--\eqref{eq:problem_subset} into a tractable three-stage pipeline comprising DT-assisted beam prediction, region-based localization, and XAI-guided beam prioritization. The detailed architecture implementing these objectives is presented in Section~IV.
 
\begin{figure}[t]
    \centering
    \vspace{-3mm}
    \includegraphics[width=9cm]{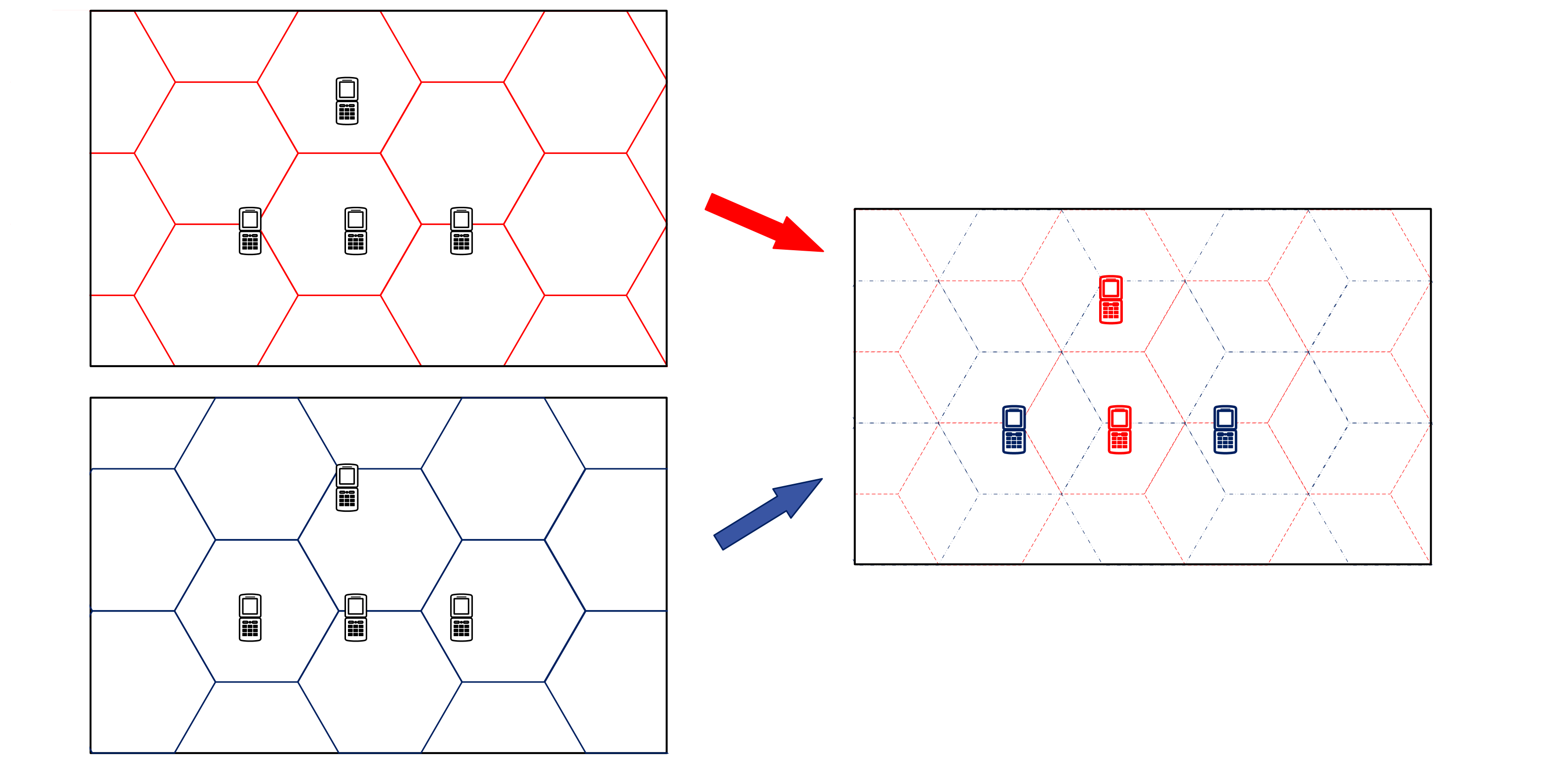}
    \caption{User-to-region assignment under the overlapping region partition.}
    \label{fig:2}
    \vspace{-5mm}
\end{figure}

\section{Region-Based User Association}
\label{sec:region_assignment}
 
The intractability of the latency-aware beam management problem
\eqref{eq:problem_obj}--\eqref{eq:problem_subset} stems chiefly from the dependence of
$\hat{\mathcal{V}}_k$ on the unknown channel $\mathbf{g}_k$ and on the fidelity of the DT.
A direct attempt to predict per-user beam subsets from DT priors would be unduly sensitive to
the inevitable mismatch term $\Delta_k$ in \eqref{eq:mismatch}, since even small errors in the
inferred user position can translate into substantial deviations between predicted and
actual dominant AoDs. To circumvent this brittleness, we discretize the coverage area into a finite
collection of \emph{regions} and lift the beam-subset design problem from the per-user level to the
per-region level. Each region is then equipped with its own beam subset, and a user is associated
with a region rather than with a precise spatial point. This indirection trades the unforgiving
requirement of exact localization for the far milder requirement of approximate region
membership, providing the structural robustness that DT-driven prediction lacks on its own.
 
\subsection{Regional Partitioning of the Coverage Area}
 
Let $\mathcal{A} \subset \mathbb{R}^2$ denote the two-dimensional ground projection of the cell
coverage area. We partition $\mathcal{A}$ using a regular hexagonal lattice of side length $a$,
chosen so that the hexagon dimension is commensurate with the angular resolution of the BS
beam codebook. The use of a hexagonal tiling, as opposed to a rectangular or Voronoi
partition, is motivated by its near-isotropic spatial coverage, its uniform center-to-vertex
distance, and its prevalence in cellular geometry; together these properties yield region
shapes whose interior points share statistically similar AoD distributions with respect to the
BS, which is precisely the property that justifies sharing a common beam subset across the
region. We denote the resulting collection of hexagonal cells, henceforth referred to as the
\emph{primary tiling}, by
\begin{equation}
\mathcal{R} \;=\; \{ \mathcal{R}_1,\, \mathcal{R}_2,\, \ldots,\, \mathcal{R}_R \},
\qquad
\mathcal{R}_r \subset \mathcal{A},
\quad
\bigcup_{r=1}^{R} \mathcal{R}_r \,=\, \mathcal{A},
\end{equation}
where adjacent hexagons share only their boundary edges so that the interiors form a disjoint
partition of $\mathcal{A}$. Each region $\mathcal{R}_r$ is identified by the planar coordinates of
its center, $\mathbf{o}_r \in \mathbb{R}^2$, with the set of all primary centers denoted
$\mathcal{O} = \{ \mathbf{o}_1, \ldots, \mathbf{o}_R \}$. Associated with $\mathcal{R}_r$ is a
region-specific beam subset
\begin{equation}
\mathcal{B}_r \;\subseteq\; \mathcal{V},
\qquad
|\mathcal{B}_r| \;\ll\; C,
\label{eq:region_beam_set}
\end{equation}
which contains the codewords identified as informative for users located anywhere in
$\mathcal{R}_r$. The construction of $\mathcal{B}_r$ from DT-generated channel data is performed
offline through an explainability-driven beam ranking procedure detailed in
Section~V; for the purposes of the present section, we treat $\mathcal{B}_r$ as a given
attribute of each region.
 
\subsection{Boundary Ambiguity and the Need for an Auxiliary Tiling}
 
Although the primary tiling cleanly partitions $\mathcal{A}$, it suffers from a well-known
artifact common to all finite-grained spatial discretizations: users whose estimated location
falls near the boundary of a hexagon are vulnerable to misclassification. Let
$\widetilde{\mathbf{p}}_k \in \mathbb{R}^2$ denote the estimated planar location of user $k$, obtained
from any of the localization modalities reported in the literature, including LiDAR sensing,
uplink-pilot-based angle/range estimation at the BS, or GPS, and assumed to satisfy
\begin{equation}
\widetilde{\mathbf{p}}_k \;=\; \mathbf{p}_k \,+\, \mathbf{e}_k,
\qquad
\|\mathbf{e}_k\| \,\leq\, \varepsilon,
\label{eq:loc_error}
\end{equation}
where $\mathbf{p}_k$ is the true position and $\varepsilon$ bounds the localization error.
Specifying the particular localization mechanism is beyond the scope of this paper. If $\mathbf{p}_k$ lies within a distance $\varepsilon$ of the edge of some hexagon, the noisy
estimate $\widetilde{\mathbf{p}}_k$ may straddle the boundary and be assigned to the wrong region.
The consequences are non-trivial: because $\mathcal{B}_r$ is tuned to the angular statistics of
$\mathcal{R}_r$, applying $\mathcal{B}_{r'}$ for a neighbor $r' \neq r$ can yield a beam subset
that omits the user's true dominant beam. In other words, while the regionalization
contains DT mismatch to local neighborhoods, it does not by itself eliminate the
sensitivity of edge-located users to localization noise.
 
To resolve this boundary ambiguity, we introduce a second hexagonal tiling
$\widetilde{\mathcal{R}}$, congruent to $\mathcal{R}$ but spatially offset, such that the regions in which
users are most vulnerable under one tiling become the regions in which they are most secure
under the other. Specifically, the auxiliary tiling is obtained by translating the primary
tiling by a shift vector
\begin{equation}
\boldsymbol{\tau} \;=\; a\,[\cos\theta_s,\, \sin\theta_s]^{\mathsf{T}},
\qquad
\theta_s \,\notin\, \left\{ \tfrac{m\pi}{3} \,:\, m \in \mathbb{Z} \right\},
\label{eq:shift}
\end{equation}
where $\|\boldsymbol{\tau}\| = a$ equals the hexagon side length and the orientation
$\theta_s$ is constrained to avoid the six directions parallel to the principal edges of the
hexagonal lattice. This constraint is essential: any shift along an edge direction would
realign the auxiliary lattice with the primary lattice and reproduce the same boundary
structure, defeating the purpose of the second tiling. Choosing $\theta_s$ off the edge axes
ensures that the centers of $\widetilde{\mathcal{R}}$ lie squarely in the interior of regions of
$\mathcal{R}$, so that every boundary of $\mathcal{R}$ is geometrically covered by an interior region
of $\widetilde{\mathcal{R}}$, and vice versa, as illustrated in Fig. 2.
The resulting auxiliary collection is
\begin{equation}
\widetilde{\mathcal{R}} \;=\; \{ \widetilde{\mathcal{R}}_1,\, \widetilde{\mathcal{R}}_2,\, \ldots,\, \widetilde{\mathcal{R}}_R \},
\qquad
\widetilde{\mathbf{c}}_r \,=\, \mathbf{c}_r + \boldsymbol{\tau},
\end{equation}
with each auxiliary region $\widetilde{\mathcal{R}}_r$ likewise endowed with its own beam subset
$\widetilde{\mathcal{B}}_r \subseteq \mathcal{V}$ obtained through the same DT-XAI procedure used
for the primary tiling.
 
\subsection{User-to-Region Association Rule}
 
Given the two tilings, every user is by construction interior to either a primary region, or an
auxiliary region. To exploit this property, we associate each user with the
\emph{nearer center} across the two tilings, on the rationale that a user closer to a region
center is deeper in that region's interior and therefore more robust to localization noise of
magnitude $\varepsilon$. Formally, define the nearest primary and auxiliary centers for user
$k$ as
\begin{align}
r_k^{\mathrm{P}}
\;&=\;
\arg\min_{r \in \{1,\ldots,R\}}
\left\| \widetilde{\mathbf{p}}_k - \mathbf{o}_r \right\|,
\label{eq:nearest_primary}\\[2pt]
r_k^{\mathrm{A}}
\;&=\;
\arg\min_{r \in \{1,\ldots,\widetilde{\mathcal{R}}\}}
\left\| \widetilde{\mathbf{p}}_k - \widetilde{\mathbf{o}}_r \right\|,
\label{eq:nearest_aux}
\end{align}
with corresponding center distances
\begin{equation}
d_k^{\mathrm{P}}
\,=\,
\left\| \widetilde{\mathbf{p}}_k - \mathbf{o}_{r_k^{\mathrm{P}}} \right\|,
\qquad
d_k^{\mathrm{A}}
\,=\,
\left\| \widetilde{\mathbf{p}}_k - \widetilde{\mathbf{o}}_{r_k^{\mathrm{A}}} \right\|.
\end{equation}
The assigned region for user $k$ is then
\begin{equation}
\Omega_k
\;=\;
\begin{cases}
\mathcal{R}_{\,r_k^{\mathrm{P}}}, & \text{if } d_k^{\mathrm{P}} \,\leq\, d_k^{\mathrm{A}},\\[4pt]
\widetilde{\mathcal{R}}_{\,r_k^{\mathrm{A}}}, & \text{otherwise},
\end{cases}
\label{eq:assignment_rule}
\end{equation}
and the corresponding beam subset to be probed by user $k$ is
\begin{equation}
\mathcal{B}(\Omega_k)
\;=\;
\begin{cases}
\mathcal{B}_{\,r_k^{\mathrm{P}}}, & \Omega_k \,=\, \mathcal{R}_{\,r_k^{\mathrm{P}}},\\[4pt]
\widetilde{\mathcal{B}}_{\,r_k^{\mathrm{A}}}, & \Omega_k \,=\, \widetilde{\mathcal{R}}_{\,r_k^{\mathrm{A}}}.
\end{cases}
\label{eq:beam_subset_selected}
\end{equation}
The geometric guarantee underpinning this rule is that, for any point $\mathbf{p} \in \mathcal{A}$,
\begin{equation}
\min\!\left( d^{\mathrm{P}}(\mathbf{p}),\, d^{\mathrm{A}}(\mathbf{p}) \right)
\;\leq\;
\frac{\sqrt{3}}{2}\,a,
\label{eq:bound}
\end{equation}
where $d^{\mathrm{P}}(\mathbf{p})$ and $d^{\mathrm{A}}(\mathbf{p})$ denote the distances from
$\mathbf{p}$ to the nearest primary and auxiliary centers, respectively, and $\tfrac{\sqrt{3}}{2}a$
is the apothem (center-to-edge distance) of the hexagon. Consequently, every user is
associated with at least one region whose center lies no farther than one apothem from the
user, ensuring that the localization-error ball of radius $\varepsilon$ around
$\widetilde{\mathbf{p}}_k$ remains well-contained in the interior of $\Omega_k$ whenever
$\varepsilon < \tfrac{\sqrt{3}}{2}a - \max(d_k^{\mathrm{P}}, d_k^{\mathrm{A}})$.
 
\subsection{Implications for Beam Training}
 
Once $\Omega_k$ is determined via \eqref{eq:assignment_rule}, the beam-selection problem
\eqref{eq:opt_beam} reduces, for user $k$, to the localized search
\begin{equation}
\hat{c}_k
\;=\;
\arg\max_{c \,:\, \mathbf{v}_c \,\in\, \mathcal{B}(\Omega_k)}
\left| \mathbf{g}_k^{\mathsf{H}} \mathbf{v}_c \right|^2,
\label{eq:region_beam_selection}
\end{equation}
which requires probing only $|\mathcal{B}(\Omega_k)|$ codewords instead of the full $C$. Since
$|\mathcal{B}(\Omega_k)| \ll C$ by design, the proposed region-based association achieves a
substantial reduction in training latency relative to exhaustive sweeping, while the dual-tiling
construction ensures that the reduction does not come at the cost of robustness for users
located near regional boundaries. The mechanism by which each $\mathcal{B}_r$ and
$\widetilde{\mathcal{B}}_r$ is constructed from the DT-generated channel knowledge, using XAI to
rank and select the most informative beams within each region, is the subject of
Section~V.

\begin{figure}[t]
    \centering
    \includegraphics[width=9cm]{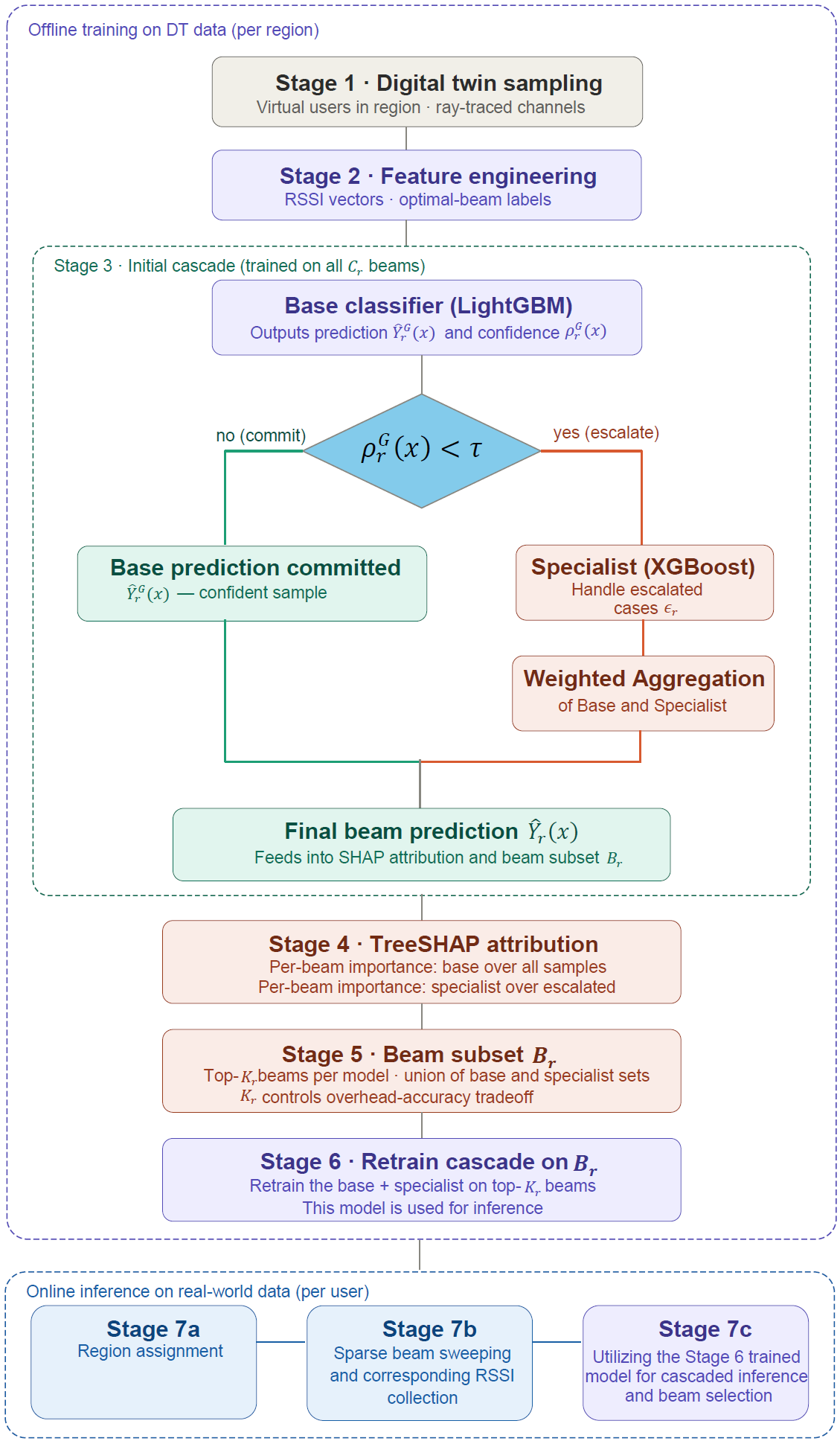}
    \caption{ The proposed stages of the offline training and online inference framework}
    \label{fig:1}
\end{figure}

\section{XAI-Guided Beam Subset Construction}
\label{sec:xai}

Having established in Section~\ref{sec:region_assignment} how each user is associated with a
region $\Omega_k \in \mathcal{R} \cup \widetilde{\mathcal{R}}$, the remaining task is to specify how the
region-specific beam subsets $\mathcal{B}_r$ and $\widetilde{\mathcal{B}}_r$ are obtained from the
DT-generated channel knowledge. The naive choice is keeping the top-ranked beams by average gain and ignores the joint statistical role that beams play when used together as input to a
predictor, and tends to retain mutually redundant codewords pointing in similar angular
directions. We instead pose subset construction as a \emph{feature selection} problem driven
by an interpretable, classifier-grounded importance measure. Concretely, for every region we
train a two-stage cascaded classifier whose task is to predict the optimal beam from a vector
of beam-power observations, and we use TreeSHAP to attribute the classifier's decisions back
to individual codewords. Beams whose attribution falls below a relevance threshold are pruned,
yielding compact subsets that retain the codewords actually responsible for reliable
prediction.

\subsection{Per-Region Dataset Generation from the DT}

Consider a fixed region $\mathcal{R}_r$ (the treatment for an auxiliary region
$\widetilde{\mathcal{R}}_r$ is identical). Using the DT of the deployment site, we sample $M_r$ virtual
user locations $\{ \mathbf{p}_{r,m} \}_{m=1}^{M_r}$ uniformly distributed within $\mathcal{R}_r$.
For each such location, the DT generates a site-specific channel realization via ray-tracing,
from which the RSSI is recorded on each beam of the full codebook $\mathcal{V}$. Stacking the
per-beam RSSI values yields the feature vector
\begin{equation}
\mathbf{x}_{r,m}
\;=\;
\left[
\mathrm{RSSI}_{r,m}^{1},\,
\mathrm{RSSI}_{r,m}^{2},\,
\ldots,\,
\mathrm{RSSI}_{r,m}^{C}
\right]^{\mathsf{T}}
\;\in\;
\mathbb{R}^{C},
\label{eq:feature_vector}
\end{equation}
where $\mathrm{RSSI}_{r,m}^{c}$ denotes the received signal strength on beam $c$ at
location $\mathbf{p}_{r,m}$. The classification label for sample $m$ is the index of the beam
yielding the highest RSSI,
\begin{equation}
y_{r,m}
\;=\;
\arg\max_{c \in \{1,\ldots,C\}} \, [\mathbf{x}_{r,m}]_c,
\label{eq:label}
\end{equation}
where $[\,\cdot\,]_c$ extracts the $c$-th entry. The resulting per-region dataset
\begin{equation}
\mathcal{D}_r \;=\; \left\{ \,(\mathbf{x}_{r,m},\, y_{r,m}) \,\right\}_{m=1}^{M_r}
\end{equation}
serves as the training corpus for the cascaded classifier associated with $\mathcal{R}_r$. We
emphasize that $\mathcal{D}_r$ is generated entirely offline from the DT and incurs no
over-the-air measurement cost.

\subsection{Cascaded Base--Specialist Classifier}

Within each region, the optimal-beam prediction task is solved by a two-stage cascade
consisting of a lightweight \emph{base} classifier and a higher-capacity \emph{specialist}
classifier, trained jointly per region but invoked sequentially at inference time. The
rationale follows the conditional-computation principle: most input samples are easy to
classify and can be resolved cheaply, whereas a residual subset of harder cases warrants a
more expressive model. Concentrating computational effort where uncertainty is high reduces
inference latency without sacrificing accuracy in the hard regime.

\paragraph{Base model.} The base $f_r^{\mathrm{G}} : \mathbb{R}^{C} \rightarrow \Delta^{C-1}$ is
realized as a gradient-boosted decision tree ensemble (LightGBM) trained on $\mathcal{D}_r$ to
output a probability distribution over candidate beam indices,
\begin{equation}
\mathbf{p}_{r}^{\mathrm{G}}(\mathbf{x})
\;=\;
f_r^{\mathrm{G}}(\mathbf{x})
\;\in\;
\Delta^{C-1},
\end{equation}
where $\Delta^{C-1}$ denotes the $(C\!-\!1)$-dimensional probability simplex. The base's
top-1 prediction $\hat{y}_{r}^{\mathrm{G}}(\mathbf{x})$ and its associated confidence $\rho_r^{\mathrm{G}}(\mathbf{x})$ are
\begin{equation}
\hat{y}_{r}^{\mathrm{G}}(\mathbf{x})
\,=\,
\arg\max_{c} \, [\mathbf{p}_r^{\mathrm{G}}(\mathbf{x})]_c,
\quad
\rho_r^{\mathrm{G}}(\mathbf{x})
\,=\,
\max_{c} \, [\mathbf{p}_r^{\mathrm{G}}(\mathbf{x})]_c.
\label{eq:base_outputs}
\end{equation}

\paragraph{Cascade routing.} A confidence threshold $\tau \in (0,1)$, fixed by validation on
held-out DT samples, determines whether the base's prediction is committed to or whether the
sample is escalated. Define the gating indicator
\begin{equation}
\mathbb{I}_r^{\mathrm{esc}}(\mathbf{x})
\;=\;
\mathbf{1}\!\left[\, \rho_r^{\mathrm{G}}(\mathbf{x}) < \tau \,\right],
\label{eq:escalation}
\end{equation}
so that $\mathbb{I}_r^{\mathrm{esc}}(\mathbf{x}) = 1$ marks $\mathbf{x}$ as a hard sample
requiring the specialist.

\paragraph{Specialist model.} The specialist
$f_r^{\mathrm{S}} : \mathbb{R}^{C} \rightarrow \Delta^{C-1}$ is realized as an XGBoost ensemble trained on the full dataset $\mathcal{D}_r$, with higher sample weights assigned to instances flagged as hard under the base model's escalation rule. This weighted training strategy emphasizes challenging samples while preserving knowledge of the overall data distribution, enabling the specialist to improve predictions in ambiguous regions of the feature space.

\paragraph{Weighted ensemble on escalated samples}
When the base is confident, its top-1 prediction is committed directly, as the sample lies in the easy regime where the base is reliable. When the sample is escalated, however, discarding the base's probability vector in favor of the specialist alone is suboptimal: although the specialist is trained well on hard cases, the base has been exposed to the full training distribution and still carries useful soft information on a portion of the escalated samples. We therefore combine the two models' probability outputs through a weighted average rather than routing to the specialist exclusively. Let $w_G, w_S \geq 0$ denote the base and specialist weights, with $w_S > w_G$ chosen to reflect the specialist's stronger competence on hard inputs. The ensemble probability vector on an escalated sample is
\begin{equation}
    \mathbf{p}^{E}_{r}(\mathbf{x}) \;=\; \frac{w_G\, \mathbf{p}^{G}_{r}(\mathbf{x}) \;+\; w_S\, \mathbf{p}^{S}_{r}(\mathbf{x})}{w_G + w_S} \;\in\; \Delta^{C-1},
    \label{eq:ensemble_prob}
\end{equation}
and the corresponding predicted beam index is
\begin{equation}
    \hat{y}^{E}_{r}(\mathbf{x}) \;=\; \arg\max_{c}\, \left[\mathbf{p}^{E}_{r}(\mathbf{x})\right]_{c}.
    \label{eq:ensemble_pred}
\end{equation}
The combined per-region predictor is therefore
\begin{equation}
    \hat{y}_{r}(\mathbf{x}) \;=\;
    \begin{cases}
        \hat{y}^{G}_{r}(\mathbf{x}), & \mathbb{I}^{\text{esc}}_{r}(\mathbf{x}) = 0, \\[4pt]
        \hat{y}^{E}_{r}(\mathbf{x}), & \mathbb{I}^{\text{esc}}_{r}(\mathbf{x}) = 1.
    \end{cases}
    \label{eq:cascade_output}
\end{equation}

\subsection{TreeSHAP-Based Importance Attribution}

Once the cascade is trained, we quantify the contribution of each beam-power feature to the
classifier's predictions using TreeSHAP \cite{lundberg2020}, a polynomial-time exact algorithm
for computing Shapley values over tree ensembles. For a classifier $f : \mathbb{R}^C \to \Delta^{C-1}$
and a sample $\mathbf{x}$, TreeSHAP returns a matrix
$\boldsymbol{\Phi}(\mathbf{x}) \in \mathbb{R}^{C \times C}$ whose entry
$\phi_{c,c'}(\mathbf{x})$ measures the marginal contribution of feature $c$ (the $c$-th beam) to
the predicted log-odds of class $c'$ (the $c'$-th beam being optimal). The global importance of
feature $c$ for region $r$ is obtained by averaging absolute attributions across all samples
and predicted classes:
\begin{equation}
I_r^{q}(c)
\;=\;
\frac{1}{M_r C}
\sum_{m=1}^{M_r}
\sum_{c'=1}^{C}
\left|
\phi_{c,c'}^{q}(\mathbf{x}_{r,m})
\right|,
\label{eq}
\end{equation}
where $q \in \{G,\mathrm{S}\}$ denotes the base or specialist model, respectively, and $\phi_{c,c'}^{q}(\mathbf{x}_{r,m})$ is the SHAP attribution of feature $c$ to class $c'$ for sample $\mathbf{x}_{r,m}$ under model $q$.

\subsection{Region Beam Subset Construction}

The importance vectors $I_r^{q}(\cdot)$, where $q \in \{\mathrm{B}, \mathrm{S}\}$, capture complementary aspects of beam informativeness: the base reflects beams useful for the easy
majority of users in $\mathcal{R}_r$, while the specialist reflects beams useful for the
ambiguous minority. Retaining only the base's top features risks pruning codewords critical
for hard cases; retaining only the specialist's top features risks pruning codewords that
anchor the easy decisions. We therefore construct the region beam subset as the \emph{union}
of top-importance features identified by both models. For each region $\mathcal{R}_r$, where $K_r $ denote the number of
top-ranked beams retained per model. Let $\pi_r^{\mathrm{G}}(\cdot)$ and $\pi_r^{\mathrm{S}}(\cdot)$
denote permutations of $\{1,\ldots,C\}$ that sort beams by $I_r^{\mathrm{G}}$ and
$I_r^{\mathrm{S}}$ in decreasing order, respectively. The top-$K_r$ beam sets selected by the
base and specialist are

\begin{align}
\mathcal{T}_r^{\mathrm{G}}
&\;=\;
\left\{ \pi_r^{\mathrm{G}}(1),\, \pi_r^{\mathrm{G}}(2),\, \ldots,\,
\pi_r^{\mathrm{G}}(K_r) \right\}, \\[4pt]
\mathcal{T}_r^{\mathrm{S}}
&\;=\;
\left\{ \pi_r^{\mathrm{S}}(1),\, \pi_r^{\mathrm{S}}(2),\, \ldots,\,
\pi_r^{\mathrm{S}}(K_r) \right\}.
\end{align}

The final region beam subset is formed by the union
\begin{equation}
\mathcal{B}_r
\;=\;
\left\{ \mathbf{v}_c \,:\, c \,\in\, \mathcal{T}_r^{\mathrm{G}} \,\cup\,
\mathcal{T}_r^{\mathrm{S}} \right\}
\;\subseteq\;
\mathcal{V},
\label{eq:final_subset}
\end{equation}
with effective cardinality
\begin{equation}
|\mathcal{B}_r|
\;=\;
\left| \mathcal{T}_r^{\mathrm{G}} \,\cup\, \mathcal{T}_r^{\mathrm{S}} \right|
\;\leq\;
2K_r
\;\ll\;
C.
\label{eq:subset_size}
\end{equation}
The parameter $K_r$ directly controls the overhead--accuracy tradeoff: a smaller $K_r$
reduces the number of beam measurements a user must perform but risks omitting codewords
that are informative for either the easy or hard prediction regime, while a larger $K_r$
improves coverage at the cost of increased sounding overhead. The union ensures that both
the easy-case and hard-case codewords are preserved; overlap between $\mathcal{T}_r^{\mathrm{G}}$
and $\mathcal{T}_r^{\mathrm{S}}$ means the true subset size is typically strictly less than
$2K_r$. The identical procedure applied to the auxiliary tiling yields
$\widetilde{\mathcal{B}}_r$ for every $\widetilde{\mathcal{R}}_r \in \widetilde{\mathcal{R}}$.

\subsection{Retraining on the Reduced Beam Subset}
\label{subsec:retraining}

The cascade described in Section~V-B was deliberately trained on all $C$ beam-power features so that the TreeSHAP analysis could identify, without bias, which codewords carry decision-relevant information for each region. We then retrain the cascade from scratch on the reduced feature space restricted to the surviving beams $\mathcal{B}_r$, yielding a deployment-ready predictor whose input dimensionality matches the sparse sweep performed online. Concretely, let $\mathbf{P}_r \in \{0,1\}^{|\mathcal{B}_r| \times C}$ be the row-selection
matrix that extracts the entries of $\mathbf{x} \in \mathbb{R}^C$ corresponding to the beams
retained in $\mathcal{B}_r$, ordered canonically by codeword index. The reduced feature vector
for sample $m$ of region $r$ is then
\begin{equation}
\widetilde{\mathbf{x}}_{r,m}
\;=\;
\mathbf{P}_r\, \mathbf{x}_{r,m}
\;\in\;
\mathbb{R}^{|\mathcal{B}_r|},
\label{eq:reduced_feature}
\end{equation}
and the corresponding pruned dataset is
\begin{equation}
\widetilde{\mathcal{D}}_r
\;=\;
\left\{\,
\left(\,\widetilde{\mathbf{x}}_{r,m},\, y_{r,m}\,\right)
\,\right\}_{m=1}^{M_r},
\end{equation}
where the labels $y_{r,m}$ remain unchanged from \eqref{eq:label} because the ground-truth
optimal beam index is independent of which features the classifier is allowed to observe.
Using $\widetilde{\mathcal{D}}_r$, we train a fresh base--specialist cascade,
\begin{equation}
\widetilde{f}_r^{\mathrm{G}}
\,:\, \mathbb{R}^{|\mathcal{B}_r|} \rightarrow \Delta^{C-1},
\qquad
\widetilde{f}_r^{\mathrm{S}}
\,:\, \mathbb{R}^{|\mathcal{B}_r|} \rightarrow \Delta^{C-1},
\label{eq:retrained_models}
\end{equation}
following the same architecture, hyperparameter family, and confidence-threshold routing rule adopted in the initial training stage, as discussed in Section~V-B.

\subsection{Inference-Time Beam Training}

At deployment, the BS stores three artifacts per region: the beam subsets
$\{\mathcal{B}_r,\widetilde{\mathcal{B}}_r\}$, the retrained cascaded classifiers
$\{(\widetilde{f}_r^{\mathrm{G}}, \widetilde{f}_r^{\mathrm{S}})\}$ and their auxiliary
counterparts, and the region-center coordinates required for assignment. When a user $k$ arrives,
the following sequence is executed.

\paragraph{Step 1: region assignment.} The user's approximate location $\widetilde{\mathbf{p}}_k$
is acquired through the localization modality described in Section~\ref{sec:region_assignment},
and the assigned region $\Omega_k$ and its corresponding subset $\mathcal{B}_r$ are
determined via the nearest-center rule \eqref{eq:assignment_rule}.

\paragraph{Step 2: sparse beam sweep.} The BS transmits beams only on the codewords contained in
$\mathcal{B}_r$, and the user reports the resulting reduced power vector
\begin{equation}
\widetilde{\mathbf{x}}_k
\;=\;
\left[
\,|\mathbf{g}_k^{\mathsf{H}} \mathbf{v}_c|^2 \,:\, \mathbf{v}_c \in \mathcal{B}_r\,
\right]^{\mathsf{T}}
\;\in\;
\mathbb{R}^{|\mathcal{B}_r|},
\label{eq:online_feature}
\end{equation}
ordered canonically to match the input convention used during retraining.

\paragraph{Step 3: cascade inference} The trained classifier associated with region $\mathcal{R}_r$ is then applied as follows: when $\widetilde{\rho}_r^{\mathrm{G}}(\widetilde{\mathbf{x}}_k) \geq \tau$, the base model's top-1 prediction is committed directly; when $\widetilde{\rho}_r^{\mathrm{G}}(\widetilde{\mathbf{x}}_k) < \tau$, the base and specialist probability vectors are instead combined under weights $w_{\mathrm{G}}$ and $w_{\mathrm{S}}$ before taking the argmax.The resulting index identifies the predicted optimal codeword $\hat{c}_k \in \{1, \ldots, C\}$, which is committed for the ensuing data-plane transmission to user $k$. Compared with exhaustive sweeping, the proposed architecture replaces $C$ beams transmissions
per user by $|\mathcal{B}(\Omega_k)| \ll C$. 

\section{Simulation Setup and Results}
\label{sec:simulation}

\begin{figure*}[t]
    \centering
    \vspace{-3mm}

    \begin{subfigure}[b]{0.32\textwidth}
        \centering
        \includegraphics[width=\linewidth,height=3.5cm]{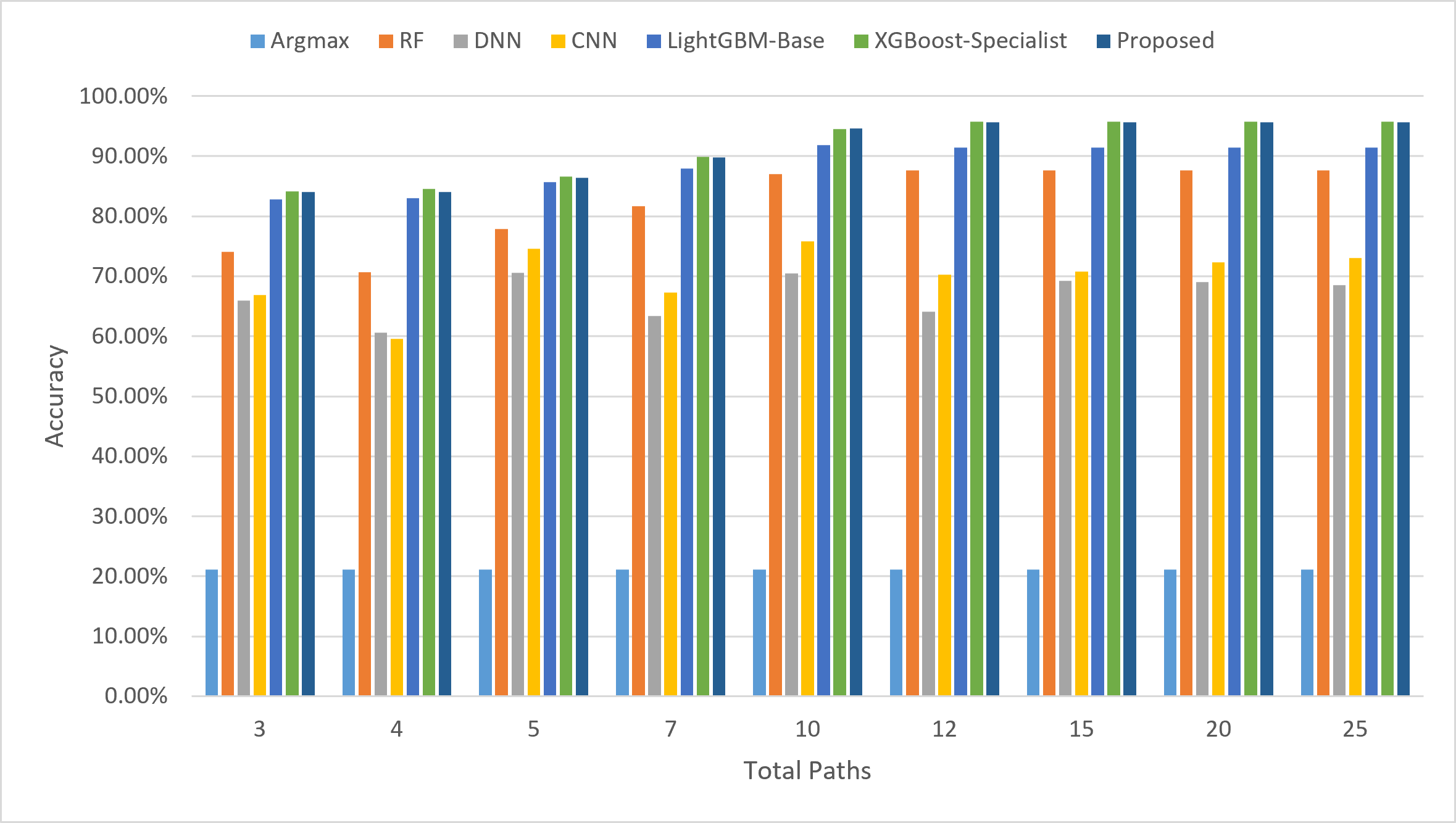}
        \caption{Top-1}       
        \label{fig:Top-1}
    \end{subfigure}
    \hfill
    \begin{subfigure}[b]{0.32\textwidth}
        \centering
        \includegraphics[width=\linewidth,height=3.5cm]{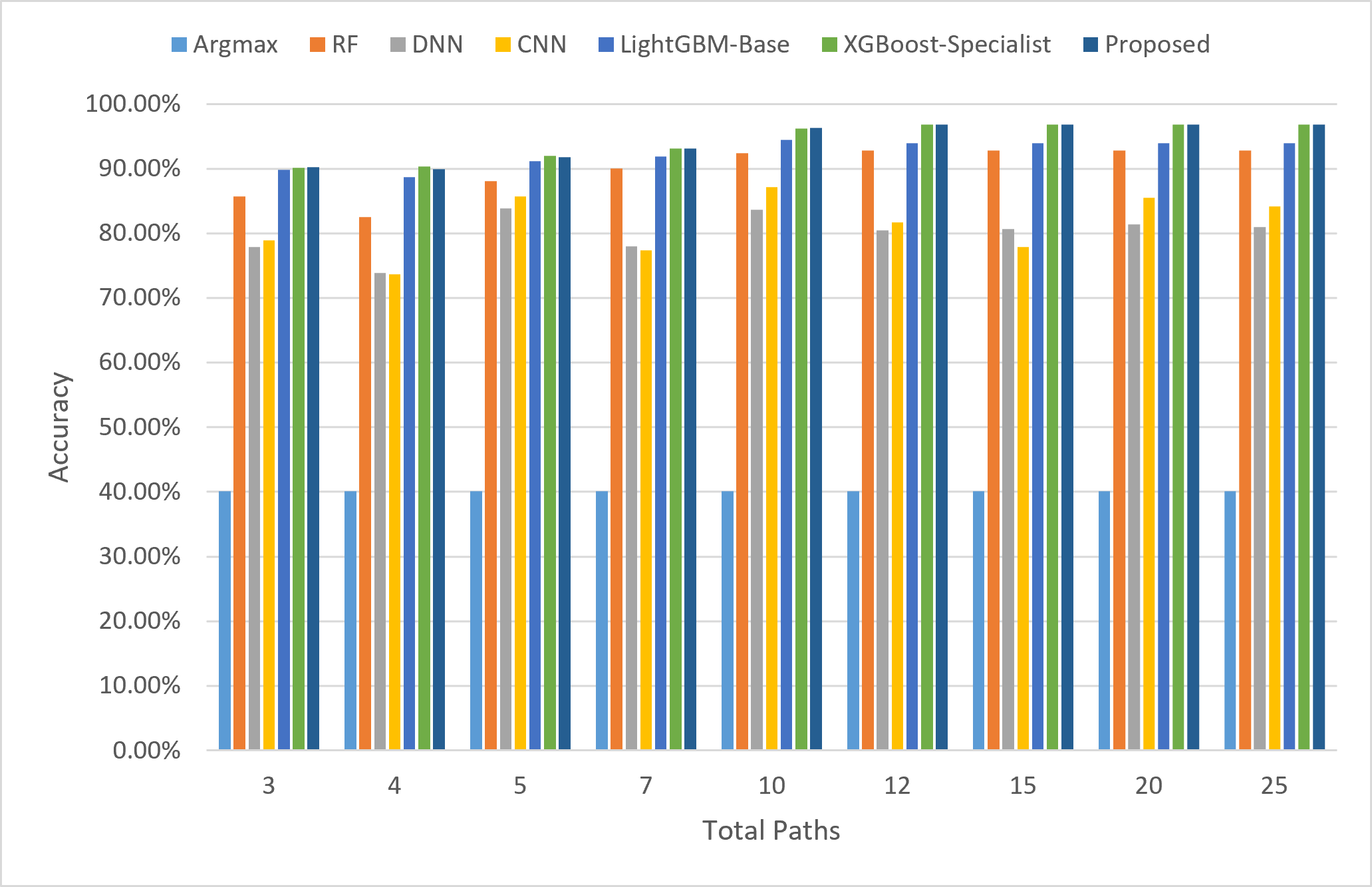}
        \caption{Top-2}
        \label{fig:Top-2}
    \end{subfigure}
    \hfill
    \begin{subfigure}[b]{0.32\textwidth}
        \centering
        \includegraphics[width=\linewidth,height=3.5cm]{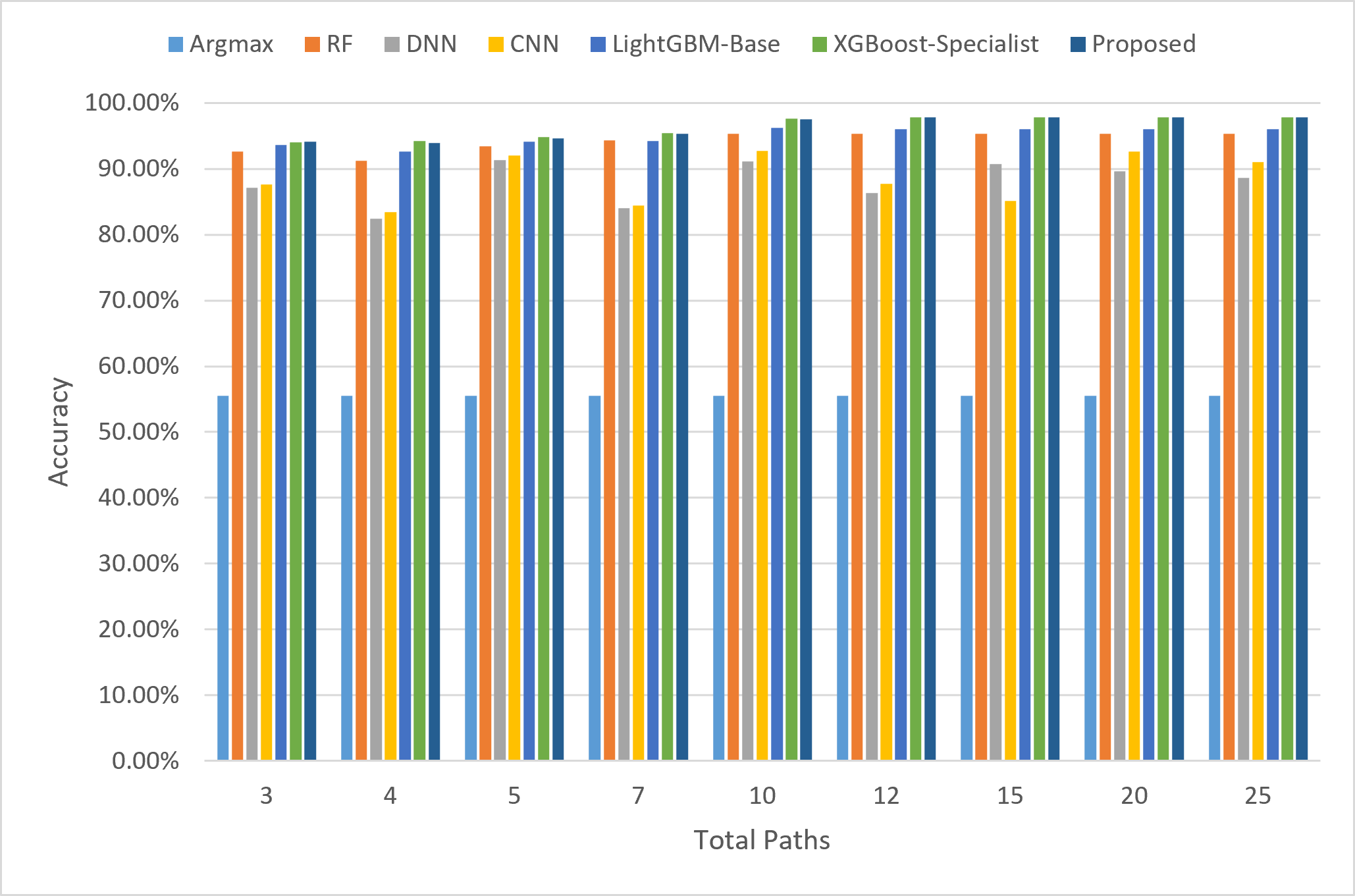}
        \caption{Top-3}
        \label{fig:Top-3}
    \end{subfigure}

    \caption{Performance comparison for Top-1, Top-2, and Top-3 beam prediction accuracy.}
    \label{fig:top123}
    \vspace{-4mm}
\end{figure*}

We evaluate the proposed framework on the DeepMIMO ray-tracing dataset using the \texttt{O1\_28} outdoor street-canyon scenario, which captures a dense urban environment populated by buildings and street infrastructure typical of cellular mmWave deployments. The base station, mounted on transmitter set TX~7, is equipped with a uniform planar array of $N = 64$ antenna elements arranged as $N_h \times N_v = 8 \times 8$ with half-wavelength inter-element spacing in both dimensions, and the array boresight is oriented toward the served user grid. The system operates at a carrier frequency $f_c = 28~\mathrm{GHz}$ with a bandwidth of $W = 500~\mathrm{MHz}$. The service area spans $600~\mathrm{m} \times 300~\mathrm{m}$ and is discretized into a uniform user grid with an inter-user spacing of approximately $0.37~\mathrm{m}$. Beam sweeping is performed using an oversampled DFT (O-DFT) codebook of size $C = 96$, providing finer angular resolution than the $N = 64$ baseline DFT.

\subsection{DT and Real-World Datasets}

To evaluate robustness under the DT mismatch term $\Delta_k$ introduced in~\eqref{eq:mismatch}, we construct a family of coupled datasets per region from the same user locations, by varying the number of propagation paths retained from the ray tracer. The \emph{real-world surrogate}, treated as ground truth, retains the full set of $25$ DeepMIMO paths per location. To emulate realistic DT imperfections, we generate eight reduced-path datasets, $\{3, 4, 5, 7, 10, 12, 15, 20\}$, plus a ninth dataset retaining all $25$ paths as an idealized upper bound where the twin exactly replicates the real channel. Discarded paths model information loss from unmodeled scatterers, calibration drift, and dynamic obstacles. The number of retained paths thus acts as a tunable DT-fidelity parameter: small values ($3$--$4$) correspond to an impoverished twin, while large values ($15$--$20$) correspond to a twin in close agreement with the real-world channel.

Throughout the simulation, classifier training is performed exclusively on a chosen DT dataset, while all reported inference results are obtained on the real-world surrogate, so that every accuracy figure reflects performance against the true channel rather than the data used for training. Each region contains $M_r \approx 7100$ user samples, partitioned $80\%/20\%$ into training and test splits at the location level: DT samples from training locations fit the per-region classifiers, while real-world surrogate samples from test locations are reserved for evaluation. No real-world surrogate data is used during training, except in the $25$-path setting, included as an idealized upper bound corresponding to a perfect DT. The coverage area is partitioned into ten overlapping hexagonal regions from the primary and auxiliary tilings of Section~\ref{sec:region_assignment} (five regions per tiling). The global codebook of $C = 96$ codewords is filtered per region prior to training, retaining only beams that appear among the top-3 strongest for at least one user in the region; the rest carry negligible energy and are excluded, leaving an effective candidate pool of roughly $40$ beams per region before any XAI-driven pruning.

\begin{figure*}[t]
    \centering
    \vspace{-3mm}

    \begin{subfigure}[b]{0.32\textwidth}
        \centering
        \includegraphics[width=\linewidth, ,height=4cm]{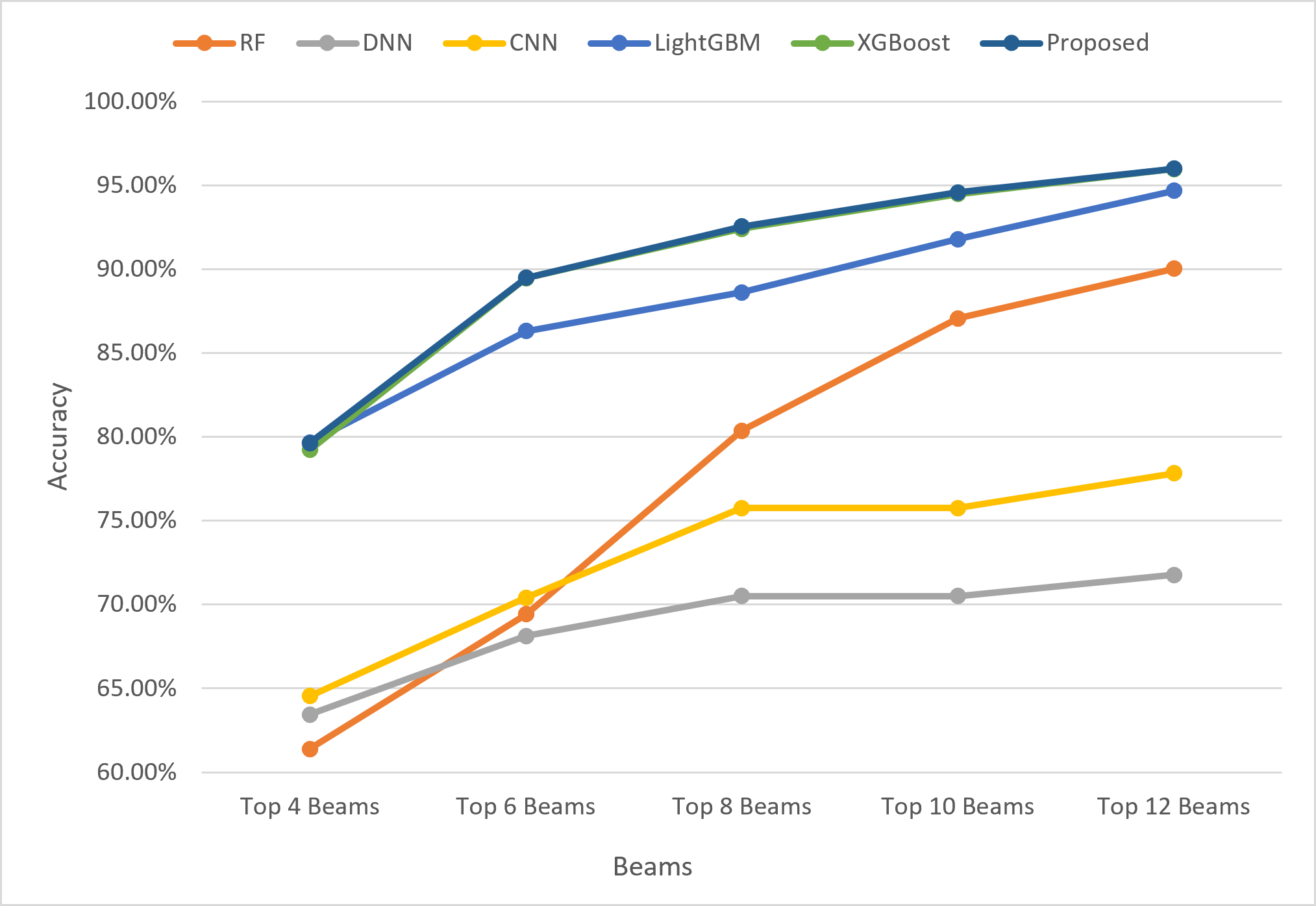}
        \caption{Top-1 Across Beams}
        \label{fig:Top-1 Across Beams}
    \end{subfigure}
    \hfill
    \begin{subfigure}[b]{0.32\textwidth}
        \centering
        \includegraphics[width=\linewidth, ,height=4cm]{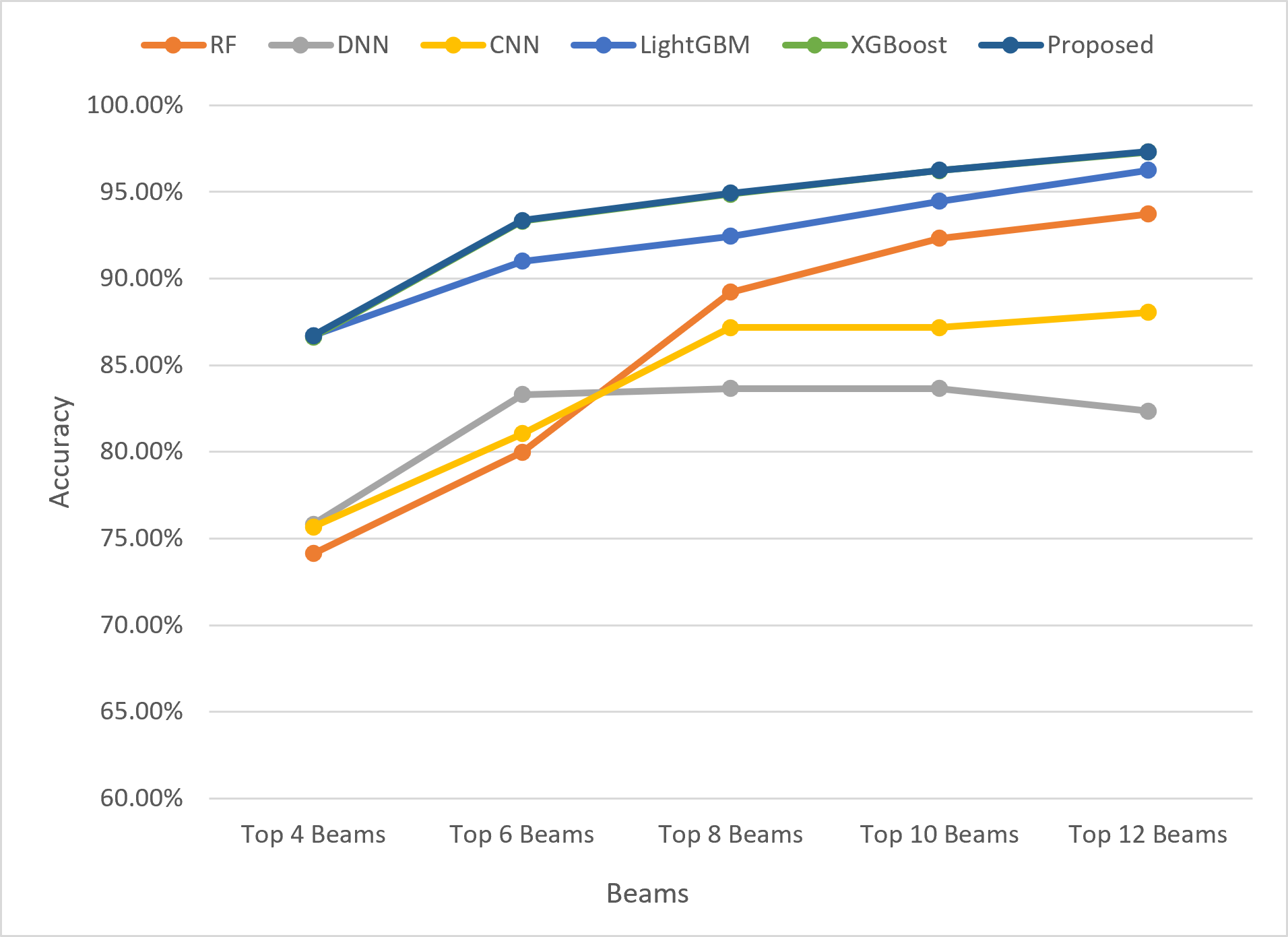}
        \caption{Top-2 Across Beams}
        \label{fig:Top-2 Across Beams}
    \end{subfigure}
    \hfill
    \begin{subfigure}[b]{0.32\textwidth}
        \centering
        \includegraphics[width=\linewidth, ,height=4cm]{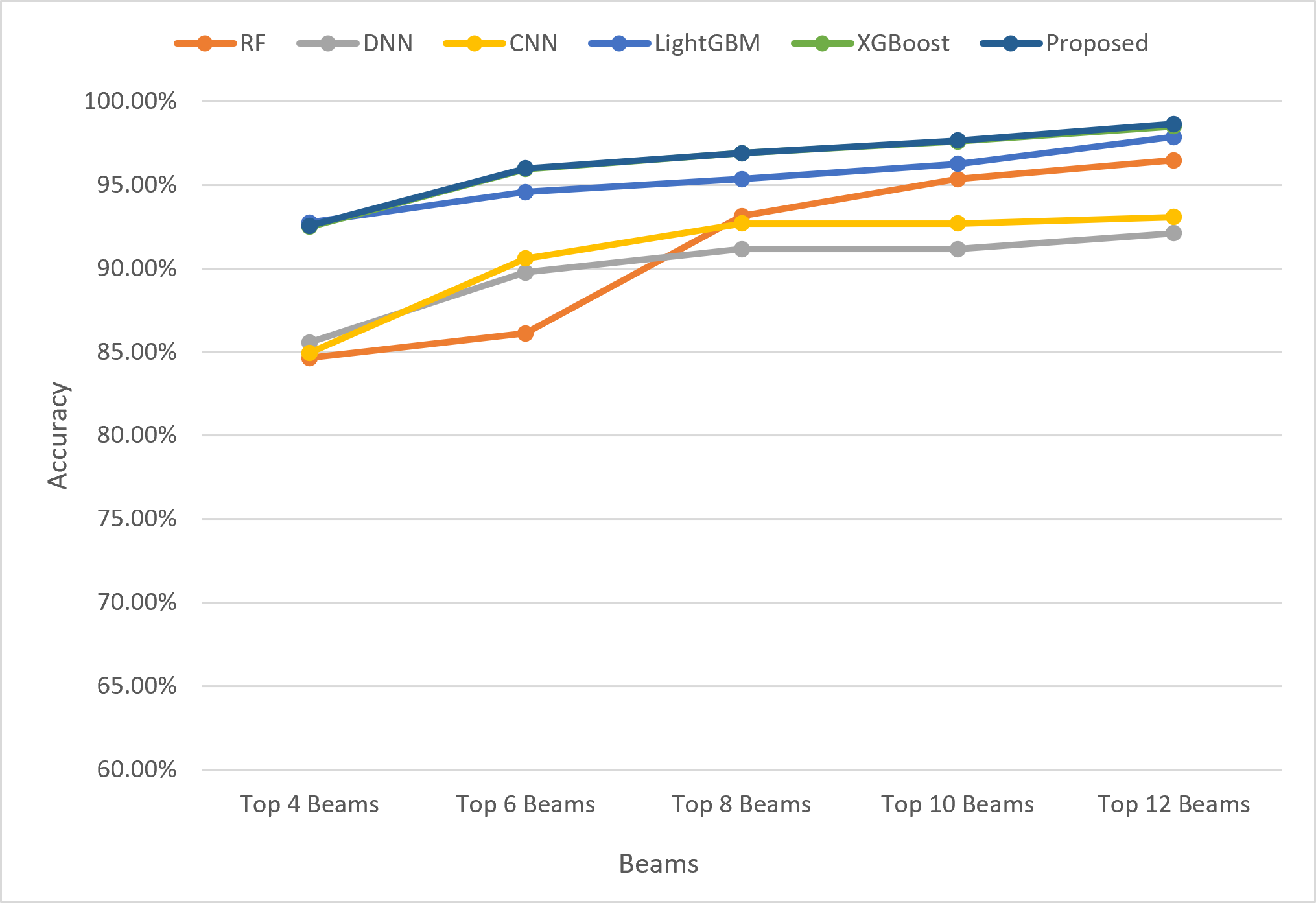}
        \caption{Top-3 Across Beams}
        \label{fig:Top-3 Across Beams}
    \end{subfigure}

    \caption{Top-$k$ beam prediction accuracy across different beam configurations.}
    \label{fig:top123beams}
    \vspace{-4mm}
\end{figure*} 

The base classifier $f_r^{\mathrm{G}}$ is implemented as a LightGBM gradient-boosted ensemble with $25$ trees, leaf count $8$, learning rate $0.2$, column and row subsampling ratios of $0.8$, and $\ell_1/\ell_2$ regularization coefficients of $0.1$ each. The specialist classifier $f_r^{\mathrm{S}}$ is implemented as an XGBoost ensemble with $100$ trees, maximum depth $8$, learning rate $0.3$, and column and row subsampling ratios of $0.8$. The cascade routing threshold is $\tau = 0.9$, the ensemble weights are $w_G = 0.5$ and $w_S = 1.5$, and the retained-beam parameter is swept over $K_r \in \{2,\, 3,\, 4,\, 5,\, 6\}$ to characterize the tradeoff between sounding overhead and prediction accuracy. Training proceeds in two stages per region. In the first stage, the initial base--specialist cascade is fit on the full $C_r$-dimensional feature vector using one-third of the regional training data, which is sufficient for stable TreeSHAP attribution while keeping the SHAP computation tractable. The resulting per-beam importances $I_r^{\mathrm{G}}$ and $I_r^{\mathrm{S}}$ feed the union-based subset construction~\eqref{eq:final_subset} and produce $\mathcal{B}_r$. In the second stage, a fresh cascade is retrained from scratch on the reduced feature space of $\mathcal{B}_r$ using the \emph{full} regional training data, yielding the deployment-ready predictor $(\widetilde{f}_r^{\mathrm{G}},\, \widetilde{f}_r^{\mathrm{S}})$.

We benchmark the proposed framework against four reference methods. The first is an \emph{Argmax} baseline that, given a feature vector $\widetilde{\mathbf{x}}_k$, predicts the beam corresponding to its largest entry without any learning; this serves as a non-parametric lower bound that captures what is achievable by trusting the DT-generated RSSI directly. The second is a \emph{Random Forest} classifier with $100$ estimators trained on the same DT data and feature space as the proposed cascade. The remaining two baselines, taken from Khan~\textit{et~al.}~\cite{khan2025explainable}, are a convolutional neural network (CNN) and a deep neural network (DNN). For consistency, all learned baselines are trained on the DT dataset and evaluated on the real-world surrogate dataset, share the same per-region candidate pool, and produce a probability distribution over the regional codewords from which the top-$k$ predictions are extracted. Prediction performance is reported using top-$k$ accuracy for $k \in \{1,\, 2,\, 3\}$, defined as the fraction of test samples for which the predicted beam $\hat{c}_k$ falls within the $k$ strongest codewords of the true codebook ordering. The top-1 metric thus reflects whether the predictor recovers the optimal beam exactly, while top-2 and top-3 quantify how close the predicted beam lies to the optimum in rank terms, with a top-2 or top-3 prediction incurring only a small beamforming-gain loss relative to the best codeword in the codebook.

\subsection{Performance Comparison with the Benchmark Schemes}
Fig.~\ref{fig:top123} reports the top-$k$ beam prediction accuracy of the proposed framework and four baselines across the nine DT-fidelity settings, with $K_r=5$, so that each region operates on at most ten codewords. For top-1 accuracy, which measures whether the predicted beam exactly matches the optimal beam, the Argmax baseline remains constant at $21.08\%$, reflecting the intrinsic difficulty of the regional beam-selection task. All learned methods substantially outperform this baseline even at three paths, showing that DT-trained models can extract useful channel structure despite limited DT fidelity. The tree-based methods exhibit a clear monotonic improvement as the number of paths increases, followed by saturation. The proposed framework reaches $95.89\%$ at saturation and outperforms XGBoost-specialist by $0.15$ percentage points. In contrast, the DNN and CNN show non-monotonic behavior: the DNN increases to $70.53\%$ at five paths, drops to $63.32\%$ at seven paths, recovers to $70.51\%$ at ten paths, and then fluctuates between approximately $64$--$70\%$. The CNN follows a similar irregular trend, remaining largely within the $60$--$76\%$ range. These fluctuations suggest that the neural models are more sensitive to distribution differences between DT-generated training data and real-world test data.

The top-2 and top-3 results follow the same overall behavior while providing a more relaxed measure of prediction quality. The proposed framework increases from $95.89\%$ top-1 to $96.88\%$ top-2 and $97.98\%$ top-3 at saturation. The relatively small $1$ percentage-point increase from top-1 to top-2 indicates that most incorrect top-1 predictions are already very close to the optimal beam. The additional $1.10$ percentage-point gain from top-2 to top-3 further shows that the remaining errors are concentrated within the immediate neighborhood of the optimal beam. The Argmax baseline similarly increases from $21.08\%$ top-1 to $40.09\%$ top-2 and $55.54\%$ top-3, confirming the intrinsic difficulty of the task without learned DT-based prediction. Among the learned models, the tree-based methods again improve monotonically with DT fidelity and then saturate, whereas the DNN and CNN remain non-monotonic. For example, the DNN top-2 accuracy reaches $83.88\%$ at five paths, falls to $77.98\%$ at seven paths, and subsequently fluctuates around $80$--$84\%$. At top-3, the proposed framework achieves $97.98\%$, exceeding XGBoost-specialist at $97.79\%$, LightGBM-base at $96.00\%$, and Random Forest at $95.29\%$. Overall, the progression from $95.89\%$ to $96.88\%$ and $97.98\%$ demonstrates that even when the optimal beam is not selected exactly, the proposed framework almost always predicts a beam among the strongest nearby codewords. With $|\mathcal{B}_r|\leq10$, this provides near-optimal beam alignment while substantially reducing the beam-sounding overhead compared with exhaustive search over the full $C=96$ codebook.

\begin{figure*}[t]
    \centering
    \begin{subfigure}[b]{0.32\textwidth}
        \centering
        \includegraphics[width=\linewidth, ,height=4cm]{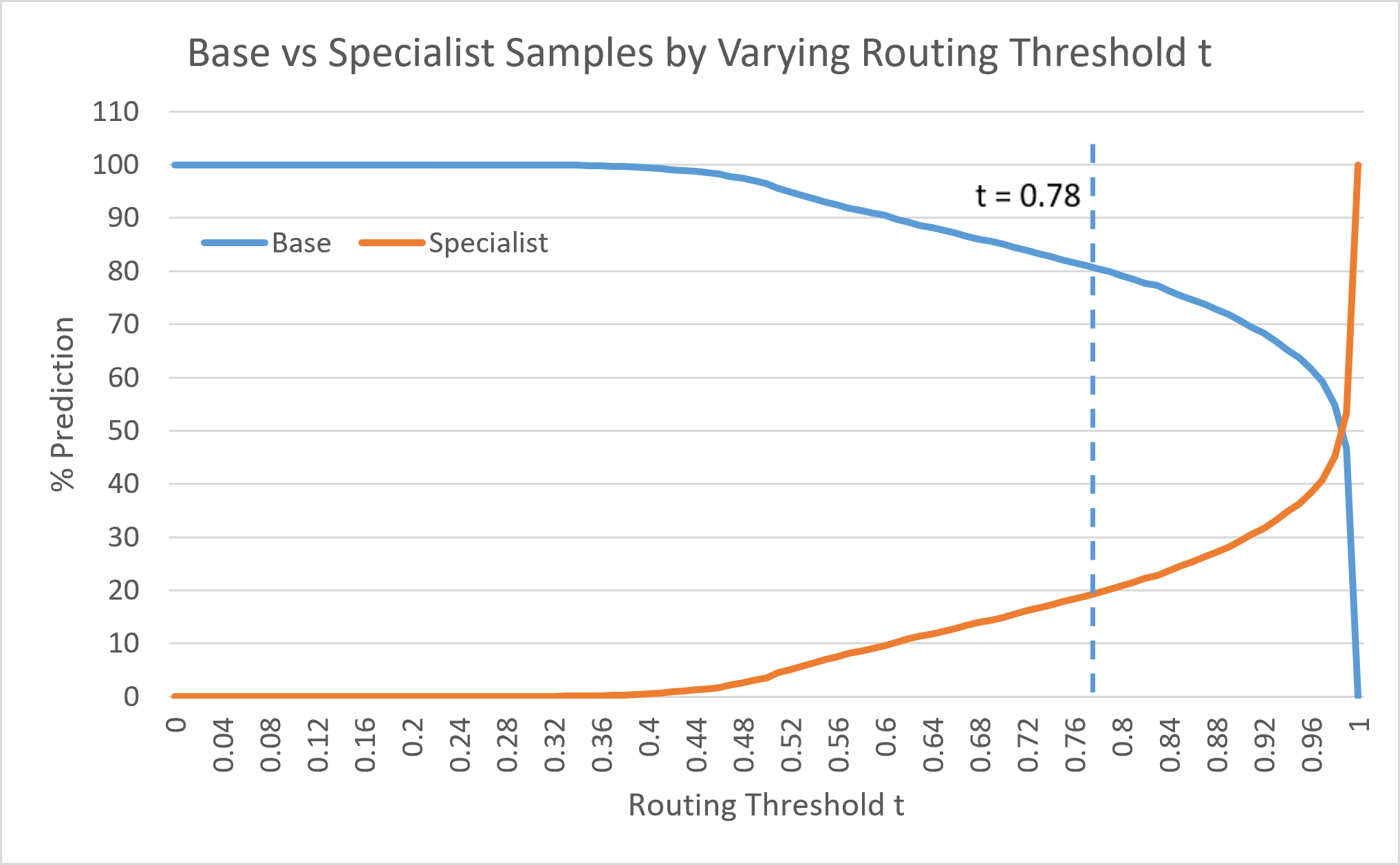}
        \caption{Sample routing}
        \label{fig:routing}
    \end{subfigure}
    \hfill
    \begin{subfigure}[b]{0.32\textwidth}
        \centering
        \includegraphics[width=\linewidth, ,height=4cm]{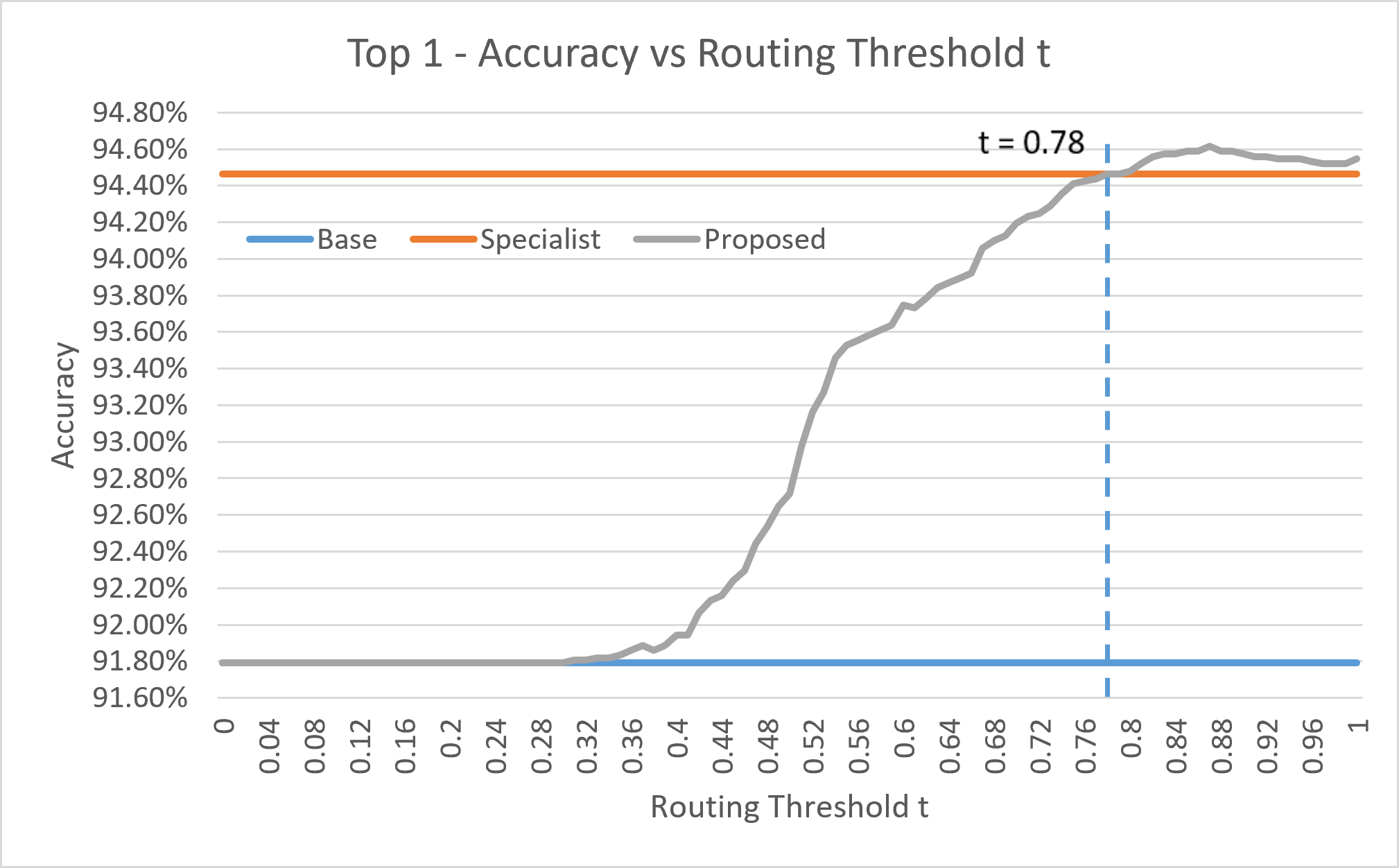}
        \caption{Top-1 accuracy}
        \label{fig:top1}
    \end{subfigure}
    \hfill
    \begin{subfigure}[b]{0.32\textwidth}
        \centering
        \includegraphics[width=\linewidth, ,height=4cm]{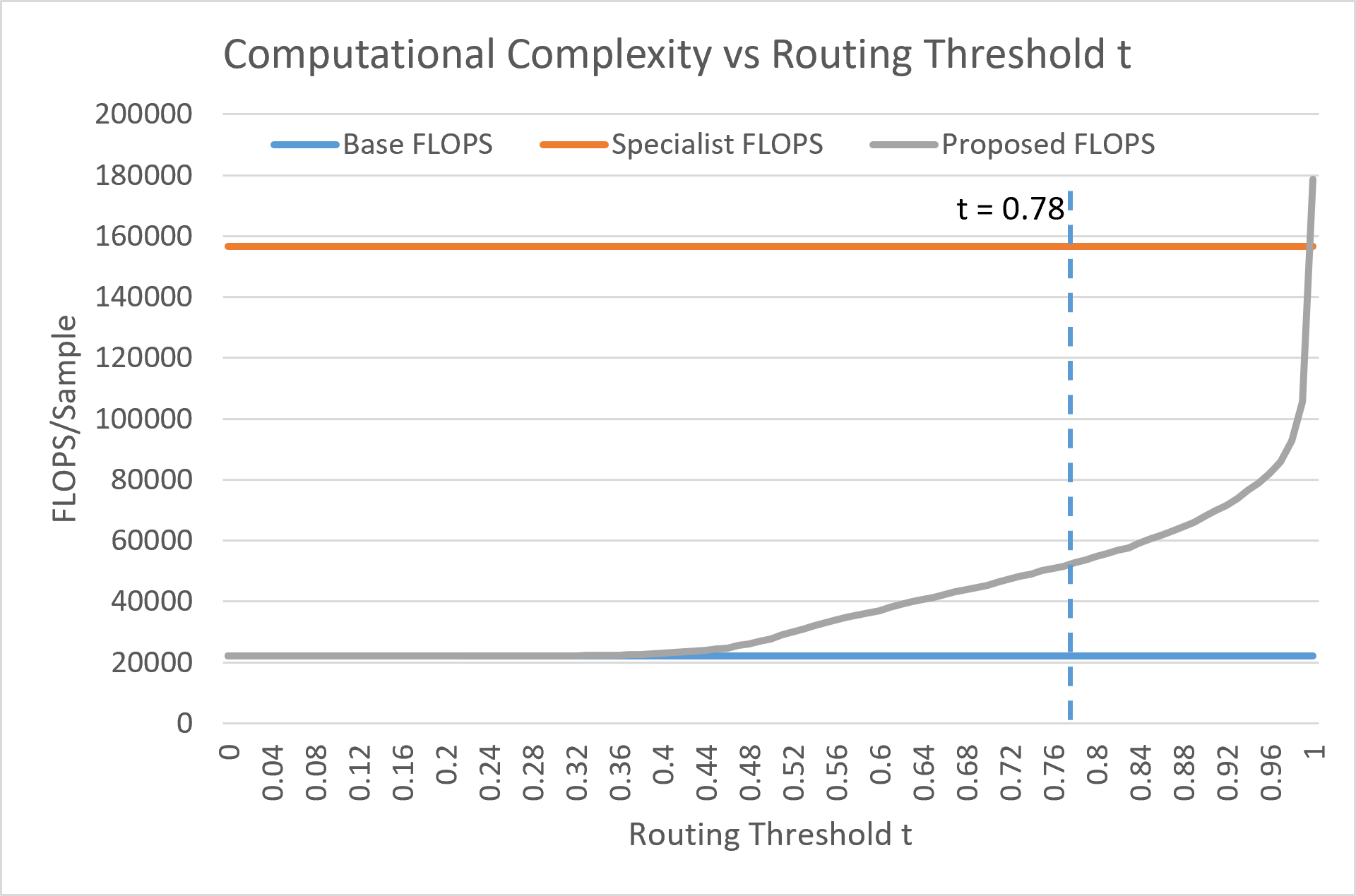}
        \caption{Computational Complexity}
        \label{fig:complexity}
    \end{subfigure}
    
    \caption{Effect of the base confidence threshold $t$ on the proposed cascade. (a)~Percentage of samples retained by the base (blue) and escalated to the specialist (orange). (b)~Top-$1$ beam-prediction accuracy of the proposed cascade against the standalone base (Base Model) and specialist (Specialist Model) baselines. (c)~ Computational complexity}
    \label{fig:threshold_sweep}
\end{figure*}

\subsection{Accuracy across different beam configurations}
Fig.~\ref{fig:top123beams} characterizes the overhead--accuracy trade-off by sweeping the retained-beam parameter over $K_r \in \{2, 3, 4, 5, 6\}$, equivalently $|\mathcal{B}_r| \in \{4, 6, 8, 10, 12\}$, at the saturation operating point of ten retained DT paths. Considering first the top-1 accuracy, the proposed framework improves monotonically with the sounding budget, climbing from $79.62\%$ at $|\mathcal{B}_r| = 4$ to $95.98\%$ at $|\mathcal{B}_r| = 12$, and dominates every alternative for all beams values. Random Forest also improves monotonically, but plateaus more slowly and reaches only $90.03\%$ at the largest budget. LightGBM-base and XGBoost-specialist likewise grow monotonically, with XGBoost-specialist tracking the proposed framework closely across the entire sweep (e.g., $95.97\%$ versus $95.98\%$ at $|\mathcal{B}_r| = 12$) while LightGBM-base trails both by two to four percentage points from $|\mathcal{B}_r| = 6$ onward. The DNN and CNN baselines behave non-monotonically, saturating between $|\mathcal{B}_r| = 8$ and $|\mathcal{B}_r| = 10$ at identical accuracies of $70.51\%$ and $75.76\%$ respectively, reflecting their sensitivity to the precise composition of the input feature space. Notably, the proposed framework already attains $92.54\%$ top-1 accuracy at $|\mathcal{B}_r| = 8$, exceeding every non-tree baseline observed at any sounding budget in this column and recovering more than $96\%$ of the accuracy achieved at $|\mathcal{B}_r| = 12$, which confirms that the union-based subset extracts most of the achievable accuracy at substantially reduced sounding cost relative to the canonical $|\mathcal{B}_r| = 10$ setting.

The top-2 accuracy follows the same trend, with the proposed framework increasing from $86.70\%$ at $|\mathcal{B}_r|=4$ to $97.32\%$ at $|\mathcal{B}_r|=12$. All tree-based methods improve monotonically, while the DNN slightly decreases from $83.65\%$ at $|\mathcal{B}_r|\in\{8,10\}$ to $82.36\%$ at $|\mathcal{B}_r|=12$, and the CNN largely saturates at $87.17\%$ for $|\mathcal{B}_r|\in\{8,10\}$, reaching only $88.05\%$ at the largest budget. At $|\mathcal{B}_r|=8$, the proposed framework achieves $94.93\%$ top-2 accuracy, indicating that the predicted beam is among the two strongest codewords in nearly $95\%$ of cases. From $|\mathcal{B}_r|=10$ onward, its accuracy remains above $96\%$, which the neural baselines do not achieve at any tested budget. Top-3 accuracy similarly approaches saturation as the sounding budget increases. The proposed framework reaches $96.91\%$ at $|\mathcal{B}_r|=8$ and increases to $97.66\%$ and $98.64\%$ at $|\mathcal{B}_r|=10$ and $12$, respectively. XGBoost-specialist follows closely at $97.60\%$ and $98.52\%$, while Random Forest remains approximately two percentage points lower. LightGBM-base also improves monotonically, reaching $97.88\%$ at $|\mathcal{B}_r|=12$. In contrast, DNN and CNN saturate earlier at $91.17\%$ and $92.70\%$, respectively, for $|\mathcal{B}_r|\in\{8,10\}$. Overall, the proposed framework achieves near-saturated top-3 accuracy with only eight codewords, while increasing the budget beyond ten provides limited additional benefit.

\subsection{Sensitivity to the base Confidence Threshold $t$}
\label{subsec:threshold_sweep}
 
To validate the proposed reliability-weighted cascade and identify a robust operating point, we sweep the base confidence threshold $t$ over $[0,1]$ in increments of $0.01$. The threshold $t$ governs the routing decision: a sample is handled by the lightweight base alone if its confidence is at least $t$, and otherwise escalated to the heavier specialist, with the two predictions fused via the reliability-weighted ensemble (Section~V). For this study, the cascade is trained on $12$-path channel realizations using the Top-$8$ feature beams, and evaluated on a more demanding $25$-path test set. Fig.~\ref{fig:routing} characterizes how the routing decision evolves with the threshold $t$. The blue curve indicates the percentage of test samples retained by the base, i.e., samples for which the base confidence is at least $t$, while the orange curve indicates the complementary fraction escalated to the specialist. For low thresholds ($t \lesssim 0.35$), the base confidence exceeds the bar on virtually all samples and no escalation occurs; the cascade therefore reduces to the base alone. As $t$ increases, progressively more samples whose confidence lies below the threshold are routed to the specialist. The escalation rate climbs gradually through the mid range ($0.35 \lesssim t \lesssim 0.85$) and rises sharply only as $t$ approaches unity. In particular, at the operating point $t^{\star} = 0.78$ only $19.83\%$ of samples are escalated to the specialist, while the remaining $80.17\%$ are handled exclusively by the base. At the limit $t = 1$, the base prediction is rejected for every sample, and all inference passes through the specialist. This confirms the efficiency of the cascade: the bulk of inference is served by the lightweight base, with the specialist invoked only for the minority of low-confidence samples.

Fig.~\ref{fig:top1} compares the Top-$1$ accuracy of the proposed cascade against the standalone base ($91.80\%$) and specialist ($94.45\%$) baselines. For $t \lesssim 0.30$, the cascade coincides with the base, since no samples are escalated. Beyond this, accuracy rises as more hard samples are corrected via reliability-weighted fusion, with a steep gain in $0.44 \leq t \leq 0.55$ marking the transition from base-dominated to specialist-informed behavior. The cascade crosses the specialist baseline near $t^{\star} = 0.78$, adopted as the operating point since it matches specialist-level accuracy while escalating fewer than one in five samples. Accuracy peaks at $94.62\%$ near $t \approx 0.86$ ($+2.82$ percentage points over the base, $+0.17$ percentage points over the specialist), achieved at an escalation fraction well below $50\%$ -- evidence that the cascade leverages the complementary strengths of both models rather than merely interpolating between them. Beyond the peak ($t \gtrsim 0.88$), accuracy decays gradually toward the specialist baseline, consistent with Fig.~\ref{fig:routing}: as $t \to 1$, the base is discarded even on samples where it is reliable, and the cascade reduces to specialist-only inference.

Fig.~\ref{fig:complexity} reports FLOPS/sample as a function of $t$. The standalone base and specialist incur constant costs of approximately $2.20 \times 10^{4}$ and $1.57 \times 10^{5}$ FLOPS/sample, respectively -- the specialist is over $7\times$ more expensive per prediction. The cascade's cost, $\mathrm{FLOPS}_\text{prop}(t) = \mathrm{FLOPS}_\text{base} + r(t)\,\mathrm{FLOPS}_\text{spec}$, increases monotonically with the escalation rate $r(t)$ from Fig.~\ref{fig:routing}: every sample incurs the base cost, and only the escalated fraction additionally incurs the specialist cost. For $t \lesssim 0.35$, escalation is negligible, and the cascade coincides with the base; as $t \to 1$, all samples are escalated, and the cascade cost exceeds the standalone specialist, since the base pass is retained on top of full specialist inference.

At $t^{\star} = 0.78$, the cascade costs approximately $5.33 \times 10^{4}$ FLOPS/sample -- only $33.9\%$ of the specialist's cost, a $2.95\times$ reduction. Together with Fig.~\ref{fig:top1}, this establishes the central efficiency claim: the cascade matches the specialist's $94.45\%$ Top-$1$ accuracy at roughly one-third of its FLOPS budget, since the base handles the $80.17\%$ of samples on which it is already confident, leaving only the remaining $19.83\%$ to incur the specialist cost. Even at the peak-accuracy setting $t \approx 0.86$ ($94.62\%$, $+0.17\%$ over the specialist), the cost is approximately $7.20 \times 10^{4}$ FLOPS -- $45.9\%$ of the specialist's cost, a $2.18\times$ reduction. The cascade thus dominates the standalone specialist on both axes across $0.30 \lesssim t \lesssim 0.90$: equal-or-higher accuracy at a fraction of the compute. This trade-off follows directly from reliability-weighted routing: the near-free base correctly disposes of most samples, so the expensive specialist is invoked only where it yields a real accuracy gain, avoiding the waste of running it on samples the base would already classify correctly. Taken together, Figs.~\ref{fig:routing}--\ref{fig:complexity} establish two key findings. First, the proposed reliability-weighted cascade matches or exceeds the accuracy of the standalone specialist across all three Top-$K$ metrics, despite the specialist being more than $7\times$ more expensive per prediction than the base. Second, the operating point is attained while routing only approximately $19.83\%$ of samples to the specialist, meaning that the base alone serves more than four out of five inference requests. These observations indicate that the proposed cascade delivers specialist-level (or better) accuracy at roughly one third of the computational footprint of always invoking the specialist.

\section{Conclusion}
\label{sec:conclusion }

This paper presented a latency-aware beam management framework for mmWave MIMO systems that combines digital-twin-based channel prediction, region-based beam localization, and explainability-driven beam pruning within a unified, cascaded classification architecture. By exploiting offline digital-twin knowledge and a base--specialist cascade with weighted ensembling, the framework resolves the majority of prediction cases efficiently while reserving heavier computation for harder cases, substantially reducing beam sounding overhead without compromising alignment accuracy. Simulation results under a digital-twin mismatch protocol demonstrate that the proposed framework achieves consistently high beam prediction accuracy, outperforming existing baselines and remaining robust across varying levels of digital-twin fidelity and beam-subset budgets. These results confirm the framework's practicality as a scalable, low-latency solution for beam management in next-generation mmWave and 6G networks.

\bibliographystyle{IEEEtran}
\bibliography{references}

\end{document}